\documentclass[preprint]{elsarticle}
\usepackage{framed,multirow}

\usepackage{amssymb}
\usepackage{latexsym}
\usepackage{placeins} % For \FloatBarrier
\usepackage{url}
\usepackage{xcolor}
\usepackage{color}
\usepackage{soul}
\usepackage{amsfonts}
\usepackage{balance}
\usepackage{tabularx}
\usepackage{multirow}
\usepackage{amsmath}
\usepackage{bm}
\usepackage{multicol}
\usepackage{booktabs}
\usepackage{rotating}
\usepackage{arydshln}
\usepackage{comment}
\usepackage{subcaption}
\usepackage{float}
\usepackage{adjustbox}
\usepackage{lineno}
\usepackage{graphicx}
\usepackage{float} %%
\usepackage{lineno}
\usepackage{hyperref}
\hypersetup{
 colorlinks=true,
 linkcolor=blue,
 filecolor=magenta, 
 urlcolor=cyan,
}

\newcommand{\bb}{\color{black}}
\newcommand{\cc}{\color{cyan}}

\usepackage{hyperref}

\definecolor{newcolor}{rgb}{.8,.349,.1}

\journal{AI Open}

\begin{document}

%\verso{Given-name Surname \textit{et~al.}}

\begin{frontmatter}

\title{ %opzione 1) AIforCOVID: an Italian multicenter study for the prediction of clinical outcomes in COVID-19 patients by applying artificial intelligence to chest-X-rays
%opzione 2) AIforCOVID: AI to predict the outcomes in patients with COVID-19 based on chest-X-rays and clinical data. An Italian multicenter study.
Multi-Scale Texture Loss for CT Denoising with GANs}
%\tnoteref{tnote1}}%
%\tnotetext[tnote1]{This is an example for title footnote coding.}

\author[1]{Francesco Di Feola\corref{cor2}} 
\author[2]{Lorenzo Tronchin}
\author[2]{Valerio Guarrasi}
\author[1,2]{Paolo Soda}
\cortext[cor2]{Corresponding author: 
email: francesco.feola@umu.se;
}

\address[1]{Department of Diagnostics and Intervention, Radiation Physics, Biomedical Engineering, Umeå University, Sweden}
\address[2]{Unit of Computer Systems and Bioinformatics, Department of Engineering, University Campus Bio-Medico of Rome, Italy}

%\received{1 May 2013}
%\finalform{10 May 2013}
%\accepted{13 May 2013}
%\availableonline{15 May 2013}
%\communicated{S. Sarkar}

\begin{abstract}
Generative Adversarial Networks (GANs) have proved as a powerful framework for denoising applications in medical imaging. However, GAN-based denoising algorithms still suffer from limitations in capturing complex relationships within the images. In this regard, the loss function plays a crucial role in guiding the image generation process, encompassing how much a synthetic image differs from a real image. To grasp highly complex and non-linear textural relationships in the training process, this work presents a novel approach to capture and embed multi-scale texture information into the loss function. Our method introduces a differentiable multi-scale texture representation of the images dynamically aggregated by a self-attention layer, thus exploiting end-to-end gradient-based optimization. We validate our approach by carrying out extensive experiments in the context of low-dose CT denoising, a challenging application that aims to enhance the quality of noisy CT scans. We utilize three publicly available datasets, including one simulated and two real datasets. The results are promising as compared to other well-established loss functions, being also consistent across three different GAN architectures. The code is available at: \url{https://github.com/TrainLaboratory/MultiScaleTextureLoss-MSTLF} 
\begin{comment} 
\end{comment}

\end{abstract}

\begin{keyword}
%% MSC codes here, in the form: \MSC code \sep code
%% or \MSC[2008] code \sep code (2000 is the default)
%\MSC 41A05\sep 41A10\sep 65D05\sep 65D17
%% Keywords
Low-dose CT (LDCT)  \sep deep learning \sep noise reduction \sep image translation, Generative Adversarial Networks \sep image synthesis 
\end{keyword}
\end{frontmatter}

\nolinenumbers

%% main text
\section{Introduction}\label{sec:Introduction}
%\linenumbers  
\label{sec:intro}
Medical imaging provides a powerful non-invasive visualization tool to reveal details about the internal human body structures. However, due to the non-idealities of the imaging processes, medical images often contain noise and artifacts, reducing the chance for an accurate diagnosis. 
Therefore, image denoising has become a crucial step in preprocessing medical images~\citep{burgos2022biomedical}, involving the inspection of a noisy image and the restoration of an approximation of the underlying clean counterpart~\citep{izadi2023image}.
%It refers to inspecting a noisy image and recovering an estimate of the underlying clean counterpart \cite{izadi2023image}.
Among different imaging modalities, Computed Tomography (CT) is of particular interest in the field of denoising. 
Basing its functioning on the absorption of X-rays, it has been proved that ionizing radiation can cause damage at different levels in biological material~\citep{brenner2007computed}. 
Due to this evidence, much effort has been put into reducing and optimizing radiation exposure under the ALARA (As Low As Reasonably Achievable) principle~\citep{bevelacqua2010practical}. 
With the increasing use of CT in many clinical applications, screenings included, the use of Low-Dose (LD) acquisition protocols has become a clinical practice (\cite{ohno2019radiation}) preferred over High-Dose (HD) protocols. 
However, if on the one side, it reduces the dose delivered to patients, on the other side, the overall quality of the reconstructed CT decreases. This motivates the investigation of image-denoising strategies that aim to obtain high-quality CT images at the lowest cost in terms of radiation. %Traditionally, it is framed as an optimization process looking for the most likely clean counterpart. 

Traditional approaches for image denoising impose explicit regularization, leveraging a priori knowledge about the noise distribution, and building a relation between relevant information and the noise present in the images. 
Some examples include dictionary learning approaches~\citep{aharon2006k} and non-local means filtering~\citep{dabov2007image}. 
These methods typically lack generalization, being able to perform well only under specific assumptions and specific hyperparameters. The advent of Deep Learning (DL) has paved the way for improved generalization capacity in denoising algorithms. 
Not needing any assumptions, DL approaches automatically learn useful features while training in an end-to-end manner, accommodating different kinds of data and they can roughly be divided into paired and unpaired methods. 
The former make use of paired datasets, i.e., they need the LD images and their respective HD counterparts, performing supervised training.
The need for paired datasets represents a strong limitation for the actual application of these algorithms. Indeed, the collection of coupled datasets is not only expensive and time-consuming, but is also unfeasible from a clinical perspective on a large scale~\citep{li2020investigation}, where efforts are made to reduce the number of examinations to the patient as much as possible. Due to the intrinsic unpaired nature of the existing datasets, \emph{unpaired deep-learning methods} are the most suitable to address CT image denoising.

Among denoising approaches, %some algorithms are based on convolutional neural networks~\cite{chen2017lowc,chen2017lowv,han2021low,nishio2017convolutional} whereas others rely on residual networks \cite{chen2017low,kang2017deep}.
%One of the most promising frameworks to tackle image denoising is 
the use of Generative Adversarial Networks (GANs) is a promising approach, being successfully employed for a variety of image-to-image translation tasks~\citep{pang2021image}.
%some others employ the paradigm of adversarial training to tackle denoising tasks with Generative Adversarial Networks (GANs) \cite{yang2018low}. 
%All these methods work in a supervised manner, i.e., 
%they need well-paired datasets in which each image must be present in both the low-dose (LD) and high-dose (HD) domains.
%This aspect
%The need for well-paired datasets represents a strong limitation for the actual application of these algorithms since the collection of coupled datasets is not only expensive and time-consuming but is also unfeasible from a clinical perspective \cite{li2020investigation}, where efforts are made to reduce the number of examinations to the patient as much as possible. Due to the intrinsic unpaired nature of the existing datasets, \emph{unpaired deep-learning methods} are the most suitable to address the problem of CT image denoising.
%GANs have been successfully employed in a variety of image-to-image translation tasks \cite{pang2021image}, representing one of the most promising frameworks to tackle image denoising.
%In this context, one of the most promising frameworks is represented by GANs \cite{goodfellow2014generative} which have been successfully employed in a variety of image-to-image translation tasks \cite{pang2021image}. 
The backbone of a GAN consists of a generator and a discriminator network. The generator learns to produce samples that fit the data distribution of the target domain while the discriminator acts as a classifier of the generated samples, attempting to distinguish between real samples and fake samples produced by the generator.
In LDCT denoising, the generator is tasked with generating realistic denoised images, whilst the discriminator is designed to differentiate between denoised images from the HDCT counterparts.
The most used backbones are Pix2Pix~\citep{isola2017image} and CycleGAN~\citep{zhu2017unpaired}. 
The former requires pairs of images in the training phase, restricting its application to paired datasets, whilst the latter relaxes this constraint with the introduction of cycle consistency, establishing itself as the milestone architecture for unpaired approaches.

To date, a variety of works have tackled paired and unpaired LDCT denoising with GANs %reaching remarkable performance, 
either modifying the baseline model architecture or introducing novel loss functions. 
Indeed, as pointed out by Huang et al.~\citep{huang2021gan}, the loss function's role is crucial, even with a more substantial impact on denoising performance than the model architecture.
While several authors used the $L_{2}$ norm (\cite{wolterink2017generative, you2018structurally, du2019visual, park2019unpaired, han2021low, yin2021unpaired, han2022perceptual, yang2023adaptive, bera2023self}), also called Mean Squared Error (MSE), it is known that such a metric produces blurred images and poorly correlate with the visual perception of the image quality (\cite{huang2021gan,wang2004image}). 
To overcome these limitations, $L_{1}$ norm is often employed as a valid alternative~\citep{ma2020low,huang2021gan, kwon2021cycle, park2022denoising, li2023multi}, but it still minimizes per-pixel errors, thus neglecting spatial and anatomical relationships between pixels.
This is why the integration of structural and semantic information into the loss function is highly recommended~\citep{elad2023image}.
%To this end, although there exists a variety of loss functions proposed in the literature \cite{izadi2023image, zhao2016loss},
To this end, although there exists a variety of loss functions  in the literature~\citep{pan2020loss,kaur2023complete}, whose review is out of the scope of this work, those based on deep perceptual similarity~\citep{yang2018low, du2019visual, li2020sacnn, yin2021unpaired, gajera2021ct, marcos2021low, li2021low, han2021low, han2022perceptual, li2022adaptive, yin2023unpaired}, structural similarity~\citep{you2018structurally, ma2020low, li2021low, li2023multi}, and Charbonnier distance~\citep{gajera2021ct, kyung2022mtd, li2022adaptive, kang2023gradient} have been adopted to a larger extent in LDCT denoising~\citep{kaur2023complete}. 
Loss functions based on deep perceptual similarity extract feature maps from a pre-trained Convolutional Neural Network (CNN)~\citep{gatys2015texture}: rather than matching pixel intensities, they minimize the distance between such deep features between the input and the corresponding synthetic image. % minimizing the difference in pixel intensities. % of aiming for a perfect match between pixel intensities. 
The underlying assumption is that the similarity between deep features may indicate the degree of perceptual similarity between the images they come from.
However, the reliability of deep features in terms of their perceptual or semantic meaning can be questioned as pointed out by~\citep{matsoukas2022makes}: deep learning models encourage the reuse of learned representations that are not necessarily associated with semantic meaning, especially when using feature extractors pre-trained on ImageNet~\citep{russakovsky2015imagenet} applied to medical domains.
The structural similarity is an alternative to $L_{1}$ and $L_{2}$ distances~\citep{elad2023image}: based on the assumption that human visual perception is highly adapted for extracting structural information from a scene, it computes the local similarity between two images as a function of luminance, contrast, and structure to assess the degradation of structural information ~\citep{wang2004image}. 
The Charbonnier distance is a smooth approximation of the Huber loss~\citep{charbonnier1994two,gajera2021ct,li2022adaptive,kang2023gradient} that is quadratic for small variations and approximates the $L_{1}$ distance for large variations. 
To the best of our knowledge, existing loss functions fail to explicitly capture texture information, a critical component in medical imaging~\citep{depeursinge2017biomedical}. While these loss functions—whether based on pixel-wise differences, structural similarity, or perceptual similarity—prioritize overall structural or perceptual fidelity, they often neglect the distinctive multi-scale texture details unique to medical images.
On these grounds, in the context of GAN-based LDCT denoising, we hereby introduce a Multi Scale Texture Loss Function (MSTLF)  that reinforces the models' ability to learn complex relationships by extracting texture information at different scales.\bb

It leverages the intrinsic multiscale nature of the Gray-Level-Cooccurence Matrix (GLCM), a handcrafted histogramming operation that counts how often different combinations of pixel intensity values occur in a specific spatial and angular relationship within an image. 
By quantifying  texture patterns and structure measuring smoothness, coarseness, and regularity~\citep{depeursinge2017biomedical,santucci20213t}, i.e., all properties that characterize an image undergoing denoising, it is a powerful tool  for texture analysis in several fields, including medical imaging. 

%To this end, it leverages the intrinsic multiscale nature of the Gray-Level-Cooccurence Matrix (GLCM). 
%Essentially the GLCM is a handcrafted histogramming operation that counts how often different combinations of pixel intensity values occur in a specific spatial and angular relationship within an image. 
%It is a powerful tool used for texture analysis in several fields, including medical imaging: it quantifies texture patterns and structure measuring smoothness, coarseness, and regularity (\cite{depeursinge2017biomedical,santucci20213t}), i.e., all properties that characterize an image undergoing denoising. 
Although the recent advances in DL have outperformed its use for classification and detection tasks, our hypothesis is that its information content can still be valuable when integrated within DL models' training. To the best of our knowledge, no one has ever tried to integrate the GLCM in a DL framework. Noticing it is non-differentiable, which makes it incompatible with gradient-based optimization, we propose a novel GLCM implementation that overcomes this limitation.
Furthermore, with our proposed MSTLF, the process of embedding handcrafted descriptors within a deep learning framework is fully enabled by the use of a self-attention mechanism that provides for the dynamic end-to-end aggregation of the multiscale information.
Experimental results on three publicly available datasets show that our approach outperforms standard loss functions and proves to be effective on different state-of-the-art GAN architectures.
%To this end, reflecting the need to define a well-suited loss function, we propose to embed texture descriptors into our denoising frameworks.
%To the best of our knowledge, no one has attempted to embed hand-crafted texture descriptors into LDCT denoising frameworks. 
\\Overall, our contributions can be summarized as follows:
\begin{itemize}
 \item We propose a novel MSTLF that leverages texture descriptors extracted at different spatial and angular scales, effectively exploiting and embedding textural information into GAN-based denoising algorithms.
 %\item \cc We introduce a novel aggregation rule based on self-attention \bb that can effectively and dynamically merge multiscale texture information.
 \item We introduce a novel aggregation rule based on self-attention that enables the model to dynamically weight and integrate information from multiple scales, resulting in a more context-sensitive and powerful representation of texture across different scales.
 \bb
 \item We propose a novel differentiable implementation of the GLCM based on a soft assignment that makes  it compatible with gradient-based optimization.
 \item We perform extensive experimentation to prove the effectiveness of the proposed approach in both paired and unpaired scenarios, testing our models on three datasets, containing both real and simulated LDCT. % of which one contains simulated LDCT images and two contain real LDCT images.
 \item We test our approach on three different GAN-based architectures, which bring out its agnostic behavior to the backbone used. %, making it suitable to be integrated in a wide range of DL frameworks.
 \item We compare our MSTLF against five state-of-the-art loss functions
    highlighting the advantages of MSTLF in terms of denoising performance.
\end{itemize}

The rest of the paper is organized as follows: section~\ref{sec:methods} presents the methods, section~\ref{sec:ExperimentalConfiguration} describes the experimental configuration, section~\ref{sec:ResultsDiscussion} presents and discusses the results, whereas section~\ref{sec:Conclusions} offers concluding remarks.

\section{Methods} \label{sec:methods}
Let $\bm{x}$ be a noisy image and $\bm{y}$ the corresponding noise-free counterpart.
The corruption by noise can be expressed as:
\begin{equation}\bm{x}= \phi(\bm{y})\label{eq}\end{equation}
where $\phi$ denotes a generic unknown degradation function. 
Image denoising aims to recover $\bm{y}$ from $\bm{x}$ by
%Hence, an approximation of the inverse function is computed: 
approximating the inverse function $\phi^{-1}$ that maps the noisy input $\bm{x}$ to its clean counterpart $\bm{\hat{y}} = \phi\bb^{-1}(\bm{x}) \approx \bm{y}$.

\begin{figure*}[h]
 \centering
 %fbox{\rule{0pt}{2in} \rule{0.9\linewidth}{0pt}}
 \includegraphics[width=11.9cm]{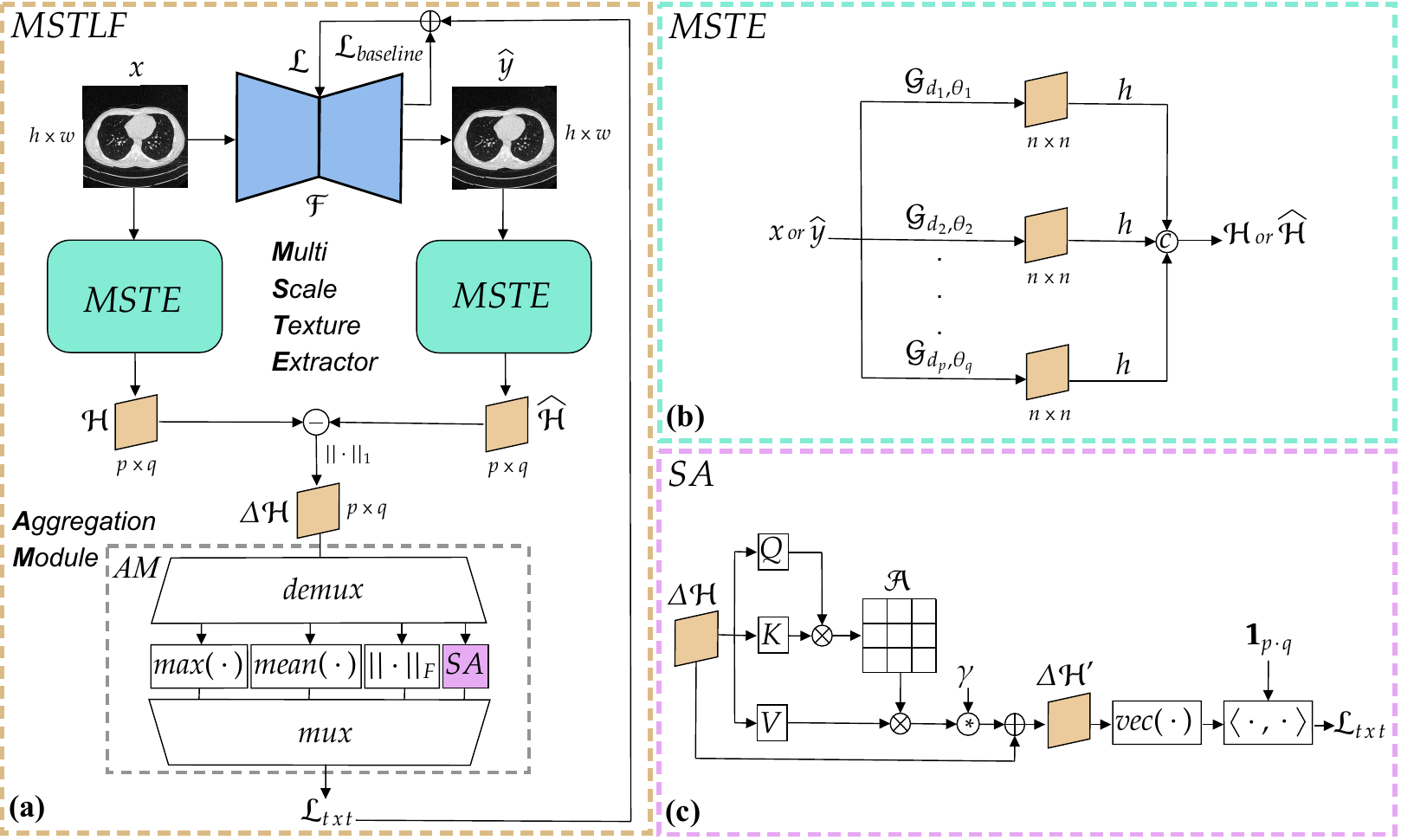}

 \caption{Overall framework of MSTLF. (a) MSTLF mainly includes two essential components, i.e., the Multi-Scale Texture Extractor (MSTE) and the Aggregation Module (AM). We denote the selection of the aggregation rule with the logic operators \emph{demux} and \emph{mux}. (b) MSTE module extracting a textural representation ($\bm{\mathcal{H}}$ or $\bm{\hat{\mathcal{H}}}$  from  the input image $\bm{x}$ or $\bm{\hat{y}}$ using a texture descriptor $h$ extracted from the GLCMs $\bm{\mathcal{G}}_{d_i, \theta_j}$ at different spatial $d_i$ and angular scales $\theta_j$). (c) Dynamic aggregation by Self-Attention mechanism (SA) that combines the extracted representation into a scalar loss function $\mathcal{L}_{txt}$.}
 \label{fig:framework}
\end{figure*}

Our MSTLF approach for GAN-based denoising applications is graphically represented in~\figurename~\ref{fig:framework}.
Panel (a) shows that MSTLF comprises two key components, integrated into the training of the denoising model $\mathcal{F}$. The first component is the Multi-Scale Texture Extractor (MSTE), detailed in panel (b), which derives a textural representation from the calculation of GLCMs at various spatial and angular scales. The second component is the aggregation module (AM) combining the textural representation into a scalar loss function $\mathcal{L}_{txt}$.
This aggregation can be performed using either static operators (maximum, average or Frobenius norm) or a dynamic approach based on self-attention (panel (c)).
The resulting $\mathcal{L}_{txt}$ is then combined with the model’s baseline loss function, $\mathcal{L}_{baseline}$, effectively embedding textural information into the training process.

%Our MSTLF approach for GAN-based denoising applications is graphically represented in~\figurename~\ref{fig:framework}.
%In panel (a), we show that MSTLF consists of two elements: a Multi-Scale Texture Extractor (MSTE) (detailed in panel (b)) that extracts a textural representation obtained from the calculation of GLCMs at different spatial and angular scales, and an aggregation module (AM) that combines the textural representation into a scalar loss function $\mathcal{L}_{txt}$ (panel (c)). 
%We implement AM using mainstream mathematical operators but also we propose a trainable approach based on self-attention Fig. \ref{fig:framework} (c). \\
In the following, we provide a rigorous mathematical description of MSTLF in section~\ref{sec:MSTLF}, while we motivate the use of texture features in section~\ref{sec:haralick}. 
In section~\ref{sec: differentability}, we solve the problem of the non-differentiability of the GLCM by introducing a novel implementation based on soft assignments, that is compatible with gradient-based optimization frameworks.

\subsection{Multi-Scale Texture Loss Function} \label{sec:MSTLF}
%GLCM is used to capture texture information which encodes the %spatial relationships (i.e., transitions) between pixel values in %an image.
%Texture information encodes the spatial relationships (i.e., transitions) %between pixel values. GLCM is a useful tool for capturing that information.
%Texture information relates to the spatial relationships (i.e., the %transitions) between pixels in an image. These transitions have properties %such as direction, scale, and order. GLCM is a 
%To capture such information, GLCM is 

% GLCM is used to count the spatial relation
Let's assume that $\bm{x} \in \mathbb{R}^{h \times w} $ and $\hat{\bm{y}} \in \mathbb{R}^{h \times w}$ are a real image a denoised image generated by a GAN model $\mathcal{F}$, where $h$ and $w$ represent their height and width.
Following ~\figurename~\ref{fig:framework} (a), both $ \bm{x} \bb$ and $\hat{\bm{y}}$ are passed to the MSTE module to extract a multiscale textural representation based on GLCMs, i.e., $\bm{\mathcal{H}}$ and $\bm{\mathcal{\hat{H}}}$.

We now describe how to extract $\bm{\mathcal{H}}$ and $\bm{\mathcal{\hat{H}}}$. 
For fixed spatial and angular offset denoted as $d$ and $\theta$, respectively, the GLCM is defined as a squared matrix $\bm{\mathcal{G}}_{d, \theta} \in \mathbb{R}^{n\times n}$, so that each element at coordinate $(i, j)$ lies in the range $[0,1]$. 
Each of its elements is computed as:
\begin{equation}
 \bm{\mathcal{G}}_{d,\theta}(i,j) = \frac{g_{d, \theta}(i, j)}{\sum_{i=1}^{n}\sum_{j=1}^{n} g_{d,\theta}(i,j)}
\end{equation}
%\begin{equation}
%\mathcal{G}_{d, \theta} = %\sum_{i=1}^{n}\sum_{j=1}^{n} %\mathcal{O}_{d,\theta}^{i,j}
%\end{equation}
where,
\begin{equation}
 g_{d,\theta}(i,j) = \sum_{u=1}^{w}\sum_{v=1}^{h} \delta[ \bm{x}(u,v)=i]\cdot \delta [ \bm{x}(\Tilde{u}, \Tilde{v})=j]
 \label{eq:glcm}
\end{equation}

s.t.
\begin{equation}
\Tilde{u} = u + d\cos\theta
\end{equation}
\begin{equation}
\Tilde{v} = v + d\sin\theta
\end{equation}
 and where $\delta$ is the Kronecker delta function, 
and $g_{d,\theta}(i,j)$ counts the occurrences of pixel value $i$ and $j$.

%Scalar statistics (e.g., contrast, dissimilarity, homogeneity), also called Haralick features, can be extracted from $\mathcal{G}_{d, \theta}$.\\

Let $D = \{d_{1}, d_{2},..., d_{p}\}$ and $\Theta = \{\theta_{1}, \theta_{2},..., \theta_{q}\}$ be the sets of spatial and angular offsets, respectively. To obtain a multiscale texture representation (~\figurename~\ref{fig:framework} (b))
% A multiscale representation is obtained starting from % the cartesian product between $D$ and $\Theta$:
we consider the cartesian product between $D$ and $\Theta$:
\begin{equation}
D \times \Theta = \{(d_{i}, \theta_{j})| d_{i} \in D, \theta_{j} \in \Theta\},
\end{equation}
%Note also that... no paired.
Let's now assume that $h$ is an operator extracting a texture descriptor from each
$\bm{\mathcal{G}}_{d_{i},\theta_{i}}$ %$\bm{\mathcal{G}}_{d,\theta}(i,j)$ 
(more details in section \ref{sec:haralick}).\\ We compute a multi-scale texture representation $\bm{\mathcal{H}} \in \mathbb{R}^{p\times  q}$ of $\bm{x}$ as:
\begin{equation}
\bm{\mathcal{H}} = \begin{bmatrix}
 h(\bm{\mathcal{G}}_{d_{1},\theta_{1}}) & ... & ... & ...& ...\\
 ... & ... & ... & ...& ...\\
 ... & ... & h(\bm{\mathcal{G}}_{d_{i},\theta_{j}}) & ...& ...\\
 ... & ... & ... & ...& ...\\
 ... & ... &... & ...& h(\bm{\mathcal{G}}_{d_{p},\theta_{q}})
\end{bmatrix}
\end{equation}
Similarly, for $\bm{\hat{y}}$ we compute $\bm{\mathcal{\hat{H}}} \in \mathbb{R}^{p\times v}$, as it is enough to replace $\bm{x}$ with $\bm{\hat{y}}$ in Eq. \ref{eq:glcm}).

Going back to~\figurename~\ref{fig:framework} (a), we then compute the error deviation $\bm{\Delta \mathcal{H}} \in \mathbb{R}^{p \times q}$ as:
\begin{equation}
 \bm{\Delta \mathcal{H}} = ||\bm{\mathcal{H}} - \bm{\mathcal{\hat{H}}}||_{1}.
\end{equation}
The aggregation module AM combines textural information into our MSTLF: 
\begin{equation}
\mathcal{L}_{txt} = AM(\bm{\Delta \mathcal{H}}).
\end{equation}
We investigate two approaches for AM that we refer to as \emph{static} and \emph{dynamic} aggregation.
In the first case, the aggregation is static because it is defined by %the fixed choice of a%
mathematical operators performing the aggregation, as follows: 
\begin{itemize}
 \item Maximum: $ \mathcal{L}_{txt} = max(\bm{\Delta \mathcal{H}})$, which leads the optimization to focus on the most discrepant texture descriptor.
 \item Average: $\mathcal{L}_{txt} = mean(\bm{\Delta \mathcal{H}})$, which leads the optimization to focus on the average discrepancy among texture descriptors.
 \item Frobenius norm: $\mathcal{L}_{txt} = ||\bm{\Delta \mathcal{H}}||_{F}= \sqrt{Tr(\bm{\Delta \mathcal{H}}\cdot\bm{\Delta \mathcal{H}}^{H})}$, similar to the average but performing a non-linear aggregation. $Tr$ is the trace operator whilst $H$ denotes the Hermitian operator.
\end{itemize}

%The former uses mathematical operators: maximum, mean, and Frobenius norm; the latter is implemented as a self-attention mechanism inspired by Han Z. et al \cite{zhang2019self}.
In the second case, the aggregation is dynamic since it enables the model to adaptively capture relationships between texture descriptors during the training of the model. 
Inspired by~\citep{zhang2019self}, this approach is implemented as a self-attention layer
(Fig. \ref{fig:framework} (c)): $\bm{\Delta \mathcal{H}}$ is passed through an attention layer which first applies $1 \times 1$ convolutions to extract keys $\bm{\mathcal{K}}$, queries $\bm{\mathcal{Q}}$ and values $\bm{\mathcal{V}}$. 
Then, the aggregation is computed as:
\begin{equation}
\mathcal{L}_{txt} = <\mathbf{1}_{p \cdot  q}, vec(\bm{\Delta \mathcal{H}}^{'})>
\end{equation}
%\begin{equation}
%\mathcal{L}_{txt} = sum(\bm{\Delta \mathcal{H}}^{'})
%\end{equation}
where,
\begin{equation}
\bm{\Delta \mathcal{H}}^{'} = \gamma\bm{\mathcal{A}}\bm{\mathcal{V}} + \bm{\Delta \mathcal{H}}
\end{equation}
s.t.
\begin{equation}
\bm{\mathcal{A}} = SoftMax(\bm{\mathcal{Q}}^{T}\bm{\mathcal{K}}).
\end{equation}
$\bm{\Delta \mathcal{H}}^{'} \in \mathbb{R}^{p\times q}$ is the output of the attention layer, and $\mathbf{1}_{p \cdot q}$ is a vector of ones of size $p \cdot q$. The term $vec(\bm{\Delta \mathcal{H}}^{'})$ reshapes $\bm{\Delta \mathcal{H}}^{'}$ into a column vector, representing the vectorized error deviation. The inner product $<\cdot,\cdot>$  computes the weighted sum of all elements in $vec(\bm{\Delta \mathcal{H}}^{'})$, with uniform weights defined by $\mathbf{1}_{p \cdot q}$. $\bm{\mathcal{A}} \in \mathbb{R}^{(p \cdot q) \times (p \cdot  q)}$ denotes the attention map, and $\gamma$ is a trainable scalar weight.
%$\bm{\Delta \mathcal{H}}^{'} \in \mathbb{R}^{p\times q}$ is the output of the attention layer, $sum(\cdot)$ is the summation operator, $\bm{\mathcal{A}} \in \mathbb{R}^{(p \cdot q) \times (p \cdot q)}$ denotes the attention map and $\gamma$ is a trainable scalar weight.

Going back to~\figurename~\ref{fig:framework} (a), after computing $\mathcal{L}_{txt}$, we combine it with the model's baseline loss function, $\mathcal{L}_{baseline}$, to form the overall loss $\mathcal{L}=\mathcal{L}_{baseline}+\mathcal{L}_{txt}$, effectively embedding multi-scale texture information into the training process.
\begin{comment}
Without loss of generality, we extend the derivation of our method to the entire image batch of size $b$:
\begin{equation}
\bm{\Xi} = [\ \xi_{1}, \xi_{2},...\xi_{b} ]\ ^{T}
\end{equation}
Then, the overall MSTLF is computed as:
\begin{equation}
\mathcal{L}_{txt} = mean(\bm{\Xi}) = \frac{1}{b} <\mathbf{1}_{b}, \bm{\Xi}>
\end{equation}
where $\mathbf{1}_{b}$ is a vector of ones of size $b$.  
\end{comment}

\subsection{Which texture descriptor?} \label{sec:haralick} 
The texture descriptor $h$, introduced in section \ref{sec:MSTLF}, can be expressed in different forms and, here, we rely on well-established Haralick features ~\citep{depeursinge2017biomedical}, namely the contrast, homogeneity, correlation, and angular second momentum, which are presented in \tablename~\ref{tab:haralicks}.
Denoting as $f(i,j)$ the term that gives a specific Haralick feature, we have for $h$
\begin{equation}
h(\bm{\mathcal{G}}_{d,\theta}) = \sum_{i=1}^{n}\sum_{j=1}^{n} f(i,j)\bm{\mathcal{G}}_{d,\theta}(i,j)
\end{equation}
%In traditional texture analysis, it is a common practice to compute a plethora of Haralick features that can provide a distinctive signature of a certain pattern (e.g., a tissue, a lesion, etc.). The use of handcrafted features to capture specific patterns within an image has been recently overcome by the advent of AI-based approaches, which offer more flexibility and generalization.
%Nonetheless, we believe that such handcrafted features can still make an informative contribution by embedding them into the optimization process of trainable models.
%These descriptors are mathematically interdependent since they can all be expressed as a function of $\bm{\mathcal{G}}_{d,\theta}$:
The same~\tablename~\ref{tab:haralicks} reports the order of magnitude of $\Delta h$, i.e., the difference between LDCT and HDCT images captured by the Haralick feature $h$ computed from 10 patients (not included in the other stages of this work) belonging to Mayo Clinic LDCT-and-Projection Data~\citep{moen2021low}, showing that contrast is the most sensitive to noise since it captures relative differences two orders of magnitude larger than the others. 
This could be expected as the contrast strictly depends on the noise in an image: indeed, the Contrast-to-Noise Ratio (CNR) of two regions $A$ and $B$ is equal to the absolute difference between their Signal-to-Noise Ratio SNR: 
\begin{equation}
CNR_{AB} = \frac{|S_{A}-S_{B}|}{\sigma_{N}} = |SNR_{A}-SNR_{B}|
\end{equation}
where $S_{A}$ and $S_{B}$ are the signals from image region $A$ and $B$, and $\sigma_{N}$ is the standard deviation of the noise.
Hence, if we assume that the noise corruption over the entire image is approximately uniform, an increase in noise corresponds to a decrease in contrast.

\begin{table*}
\centering
\begin{tabular}{|c|c|c|}
\hline
$\bm{h}$ & $\bm{f(i,j)}$ & \textbf{Magnitude of} \bm{$\Delta h$} \\
\hline
Contrast & $(i-j)^{2}$ & $\sim10^{2}$ \\

Homogeneity & $\frac{1}{1-(i-j)^{2}}$ & $\sim10^{-2}$ \\
Correlation & $\frac{(i-\mu_{i})(j-\mu_{j})}{\sqrt{\sigma_{i}^{2}\sigma_{j}^2}}$ & $\sim10^{-3}$ \\
Angular second moment & $\bm{\mathcal{G}}_{d,\theta}(i,j)$ & $\sim10^{-2}$ \\

\hline
\end{tabular}
%qui grafico
\caption{Definition of 4 well-established Haralick features and order of magnitude of the difference captured ($\Delta h$) between LDCT images and HDCT images computed from 10 patients (not included in the other stages of this work) belonging to Mayo Clinic LDCT-and-Projection Data~\citep{moen2021low}. The larger the value of  $\Delta h$ the higher the sensitivity to noise.}
\label{tab:haralicks}
\end{table*}

%--------------------------------------------------------------------------
\subsection{How to make the GLCM differentiable} \label{sec: differentability}
Differentiability is a key feature of any deep-learning framework. Whenever a new loss function or a regularization term is introduced, its implementation must be differentiable, i.e., it must be compatible with gradient-based optimization, enabling the backpropagation of the gradient throughout the entire model. 
However, the GLCM is not differentiable by its very nature, since it is essentially a histogramming operation and hence it does not align with gradient-based optimization frameworks. 
For this reason, we propose a soft approximation of the GLCM which makes it a continuous and differentiable function.

To implement a differentiable GLCM, we employ a Gaussian soft assignment function for each pixel value to a set of predefined bins:
\begin{equation}
 a_k(\bm{x}(u,v)) = e^{-\frac{(\bm{x}(u,v) - b_k)^2}{2 \sigma^2}} \label{eq:soft}
\end{equation}
where $\bm{x}(u,v)$ is the pixel value in position $(u, v)$, $b_k$ is the $k^{th}$ bin value and $\sigma$ is a hyperparameter denoting the standard deviation of the Gaussian assignment function.
The latter is chosen due to its smooth bell-shaped curve, which helps in assigning weights to neighboring bins, ensuring a soft transition between them.
Furthermore, $\sigma$ controls the spread of the Gaussian: a lower $\sigma$ produces sharper assignments (closer to hard binning), while a higher $\sigma$ makes it softer, spreading the value across multiple bins. 
In our experiments, $\sigma$ is set to 0.5 which results in a negligible error in the calculation of the soft GLCM of the order of $\sim10^{-11}$.
%The flexibility in choosing $\sigma$ allows for adaptability to various textures and noise levels.
To ensure that the weights sum up to one, preserving the nature of probability, we normalize the soft assignment for each pixel with respect to all bins:
\begin{equation}
 a'(\bm{x}(u,v)) = \frac{a_k}{\sum_{k=1}^n a_k(\bm{x}(u,v))}
\end{equation}

Given two images (original $\bm{x}$ and the corresponding spatially shifted $\bm{x_s}$ by the specified distance $d$ and angle $\theta$) and their soft assignments $a'(\bm{x}(u,v))$, the soft GLCM is obtained using the outer product:
\begin{equation}
\bm{\mathcal{G}}_{d,\theta} = \sum_{v=1}^h \sum_{u=1}^w a'(\bm{x}(u,v)) \times a'(\bm{x_{s}} (u,v))
\end{equation}
where the outer product $\times$ ensures that all possible combinations of pixel intensity relationships are considered, providing a comprehensive texture descriptor.
It is worth noting that the entire process consists of differentiable operations, from Gaussian soft assignment to normalization and the outer product followed by summation, making the soft GLCM a differentiable function with respect to the input image $\bm{x}$.
Furthermore, this method generalizes to any number of gray levels and can handle images of any bit depth without binning or down-scaling.
Similarly, we compute $\bm{\mathcal{G}}_{d,\theta}$ for $\bm{\hat{y}}$, as it is enough to replace $\bm{x}$ with $\bm{\hat{y}}$ in Eq.~\ref{eq:soft}.
%--------------------------------------------------------------------------
\subsubsection{Computational Cost Analysis}

Given that the Soft GLCM is a differentiable alternative to the traditional GLCM, it is crucial to analyze and compare the computational costs of these two methods. 

The computational process of the traditional GLCM involves iterating through each pixel in the image and updating a co-occurrence count in the GLCM matrix.
If the total number of pixels in the image is denoted as $N = h \cdot w$ and the size of the GLCM matrix as $n \times n$, the computation for each pixel pair is a constant time operation, $\mathcal{O}(1)$. 
Thus, the overall computational complexity of the traditional GLCM method scales linearly with the number of pixels, resulting in a complexity of $\mathcal{O}(N)$.

In contrast, the Soft GLCM incorporates a Gaussian soft assignment for each pixel value against a set of predefined bins, followed by the construction of the GLCM using these soft assignments. 
This procedure involves a two-fold computational process for each pixel: the soft assignment and the GLCM construction.
In the soft assignment, each pixel value is subjected to a Gaussian computation against each bin, followed by a normalization step, incurring a complexity of $\mathcal{O}(n)$ per pixel. 
The construction of the GLCM from these soft assignments involves calculating the outer product of the soft assignment vectors of each pixel pair, resulting in a computational complexity of $\mathcal{O}(n^2)$ for each pixel.
Aggregating these computations, the Soft GLCM presents a total computational complexity of $\mathcal{O}(N \cdot n^2)$, which is quadratic in the number of gray levels and linear in the number of pixels.
We note that for images with high resolution or high bit depth, the Soft GLCM becomes significantly more computationally expensive compared to the traditional GLCM, but comes with the benefit of differentiability, which is crucial for gradient-based optimization in deep learning frameworks. 
This allows its integration into the deep learning framework, which enables leveraging parallel GPU tensor computations and batch-level processing, thus mitigating the computational overhead to a large extent.

\section{Experimental Configuration}
\label{sec:ExperimentalConfiguration}
This section describes the deep models used, the competitors, the data used, the implementation details, and the evaluation metrics. 
Let us remember that $\bm{x} \in X$ is a LDCT image, whilst $\bm{y} \in Y$ is an HDCT image.
To show the effectiveness of our method we validate our approach on three of the most widely used generative architectures~\citep{pang2021image} in the image translation landscape: Pix2Pix ~\citep{isola2017image}, CycleGAN~\citep{zhu2017unpaired}, and UNIT~\citep{liu2017unsupervised}. The interested reader may refer to section 1 in the supplementary material for a detailed description of the three architectures.

%We use the default values for the hyperparameters, which are $\lambda_{0}=10$, $\lambda_{1}=\lambda_{3}=0.1$ and $\lambda_{2}=\lambda_{4}=100$.

%\begin{table*}[h]
%\begin{tabular}{|c|c|}
%\hline

%\hline
%Pix2Pix & $\mathcal{L}(G,D)=\mathcal{L}_{cGAN}(G, D)+\lambda\mathcal{L}_{1}(G)$ \\
%\hline
%CycleGAN & $\mathcal{L}(G,F,D_{X},D_{Y})=\mathcal{L}_{GAN} (G, D_{Y})+\mathcal{L}_{GAN}(F, %D_{X})+\lambda\mathcal{L}_{cyc}(G, F)$ \\
%\hline
%UNIT & \makecell{$\mathcal{L}(E_{1},E_{2}, G_{1}, G_{2}, 
%D_{1}, D_{2})= \mathcal{L}_{VAE_{1}}(E_{1},G_{1})+ \mathcal{L}_{GAN_{1}}(E_{2},G_{1}, D_{1})+
%\mathcal{L}_{CC_{1}}(E_{1},G_{1}, E_{2}, G_{2})+$\\
%$\mathcal{L}_{VAE_{2}}(E_{2},G_{2})+ \mathcal{L}_{GAN_{2}} (E_{1},G_{2}, D_{2})+
%\mathcal{L}_{CC_{2}}(E_{2},G_{2}, E_{1}, G_{1})$}\\
%\hline
%\end{tabular}
%\caption{Objective functions of State-Of-The-Art generative models.}
%\label{tab:losses}
%\end{table*}
\begin{table}[h]
\centering
\begin{tabular}{l|c|c}
\toprule
%\multicolumn{2}{c|}{OBJECTIVE FUNCTION}
&\textbf{Experiment} & \textbf{Loss function} \\
\toprule
\multirow{5}{5em}{Competitors}& Baseline & $\mathcal{L}$ \\
\cline{2-3}
&VGG-16 & $\mathcal{L}+\mathcal{L}_{VGG}$\\
\cline{2-3}
&AE-CT & $\mathcal{L}+\mathcal{L}_{AE}$\\
\cline{2-3}
&SSIM-L & $\mathcal{L}+\mathcal{L}_{ssim}$\\
\cline{2-3}
&EDGE & $\mathcal{L}+\mathcal{L}_{\nabla^{2}}$\\
\midrule
\multirow{5}{5em}{Our approach}&MSTLF-max & $\mathcal{L}+\mathcal{L}_{txt}^{max}$\\
\cline{2-3}
&MSTLF-average & $\mathcal{L}+\mathcal{L}_{txt}^{avg}$\\
\cline{2-3}
&MSTLF-Frobenius & $\mathcal{L}+\mathcal{L}_{txt}^{Frob}$\\
\cline{2-3}
&MSTLF-attention & $\mathcal{L}+\mathcal{L}_{txt}^{att}$\\
\bottomrule
\end{tabular}
\caption{List of the proposed loss function configurations. For Pix2Pix, CycleGAN, and UNIT, the baseline loss function $\mathcal{L}$ is implemented as reported in section 1 in the supplementary material.}
\label{tab:losses}
\end{table}

%--------------------------------------------------------------------------

\subsection{Competitors} \label{sec:Competitors}
To rigorously compare our method against established approaches, we include as competitors five  GAN loss terms adopted to a larger extent in LDCT denoising, which are already introduced in section~\ref{sec:intro}. 
%Furthermore, we design the experiments such that variations in performance are only attributed to variations in the loss function used. 
Furthermore, we  design the experiments to ensure that any variations in performance can be solely attributed to differences in the loss function employed.
As detailed in \tablename~\ref{tab:losses}, we evaluated a total of nine configurations for each backbone: five correspond to the loss functions used as  competitors, and the other four are different implementations of our MSTLF that vary in the type of aggregation used to combine the texture descriptors.

The first configuration for each model architecture is the baseline implemented in its original version, which we report in the supplementary material.
The second is the perceptual loss function $\mathcal{L}_{VGG-16}$ that is computed starting from the deep features extracted from the VGG-16 network~\citep{simonyan2014very}, pre-trained on ImageNet~\citep{russakovsky2015imagenet}.

This loss term, originally introduced in~\citep{gatys2015texture}, stems from observing that the similarity of features obtained from CNNs may indicate the level of semantic similarity between the images they come from providing an additional informative contribution compared to the traditional pixel-wise term~\citep{pan2020loss}.
Given a generic image $\bm{I}$ and its approximation $\bm{\Tilde{I}}$, $\mathcal{L}_{VGG}$ is computed as~\citep{johnson2016perceptual}:
\begin{equation}
\mathcal{L}_{VGG} = \frac{1}{W \cdot H \cdot D}\sum_{i=1}^{5}||\bm{\phi_{i}}(\bm{\Tilde{I}})-\bm{\phi_{i}}(\bm{I})||_{2}^2
\end{equation}
where $\bm{\phi_{i}}$ is the $i^{th}$ pooling layer of the VGG-16 and $W$, $H$, $D$ are its width, height, and depth, respectively.

As mentioned in section~\ref{sec:intro}, deep feature extractors, pre-trained on ImageNet, can produce sub-optimal features when used in domains other than the natural image domain ~\citep{matsoukas2022makes}. 
For this reason, we implement a deep feature extractor trained on CT images~\citep{han2021low,li2020sacnn, han2022perceptual}. 
We adopt the self-supervised approach proposed by~\citep{li2020sacnn} that trains an auto-encoder network to extract an encoding from the input CT image that is then used to reconstruct an output CT image as close as possible to the input. 
We trained the auto-encoder with 8 patients (not included in the other stages of this work) for a total of 2824 images; among them, 4 belong to the
Mayo Clinic LDCT-and-Projection Data~\citep{moen2021low} and the remaining 4 belong to the Lung Image Database Consortium and Image Database Resource Initiative (LIDC-IDRI) ~\citep{armato2011lung}. 
By utilizing data from two distinct datasets, we aimed to enhance the heterogeneity of the learned data distribution.
We then use the encoded representation to compute the perceptual loss $\mathcal{L}_{AE}$ (third row in \tablename~\ref{tab:losses}):
\begin{equation}
\mathcal{L}_{AE} = \frac{||\bm{e}(\bm{\Tilde{I}})-\bm{e}(\bm{I})||_{2}^2}{W \cdot H \cdot D}
\end{equation}
where $\bm{e}$ is the encoded representation and $W$, $H$, $D$ are its width, height, and depth, respectively.

The fourth loss function is based on structural similarity~\citep{wang2004image}. 
$\mathcal{L}_{ssim}$, proposed as a more sophisticated and robust distance than the traditional $L_{1}$ and $L_{2}$ distances~\citep{li2021low, li2023multi,you2018structurally}, is computed as:
\begin{equation}
\mathcal{L}_{ssim} = 1-SSIM(\bm{\Tilde{I}}, \bm{I})
\end{equation}
where $SSIM$ is the structural similarity index, described in section \ref{sub:metrics}.

To overcome the limitations posed by $L_{1}$ and $L_{2}$ distances, some authors propose the use of the Charbonnier distance which is a smooth approximation of the Huber loss ~\citep{charbonnier1994two,gajera2021ct,li2022adaptive,kang2023gradient} that is quadratic for small variations and approximates the $L_{1}$ distance for large variations. 

As fifth competitor, named  edge loss, we use the Charbonnier distance $\mathcal{L}_{\nabla^{2}}$~\citep{kyung2022mtd, zamir2021multi}, which   minimizes the distance between Laplacians instead of pixel-intensities.
%We use the Charbonnier distance  \ggr to compute the fifth competitor \bb called edge loss $\mathcal{L}_{\nabla^{2}}$ (\cite{kyung2022mtd, zamir2021multi}) that minimizes the distance between Laplacians instead of pixel-intensities. 
It is given by:
\begin{equation}
\mathcal{L}_{\nabla^{2}} = \sqrt{||\nabla^{2}(\bm{\Tilde{I}})-\nabla^{2}(\bm{I})||^{2}+\epsilon^{2}}
\end{equation}
where $\nabla^{2}$ denotes the Laplacian operator and $\epsilon^{2}$ is a constant empirically set to $10^{-3}$.
It is also interesting to note that $\mathcal{L}_{\nabla^{2}}$ catches a gradient-based quantity that maps edge information in the image, i.e., a feature different from the textural data we represent in our approach.

\begin{figure*}[t]
 \centering
 %fbox{\rule{0pt}{2in} \rule{0.9\linewidth}{0pt}}
 \includegraphics[width=1\linewidth]{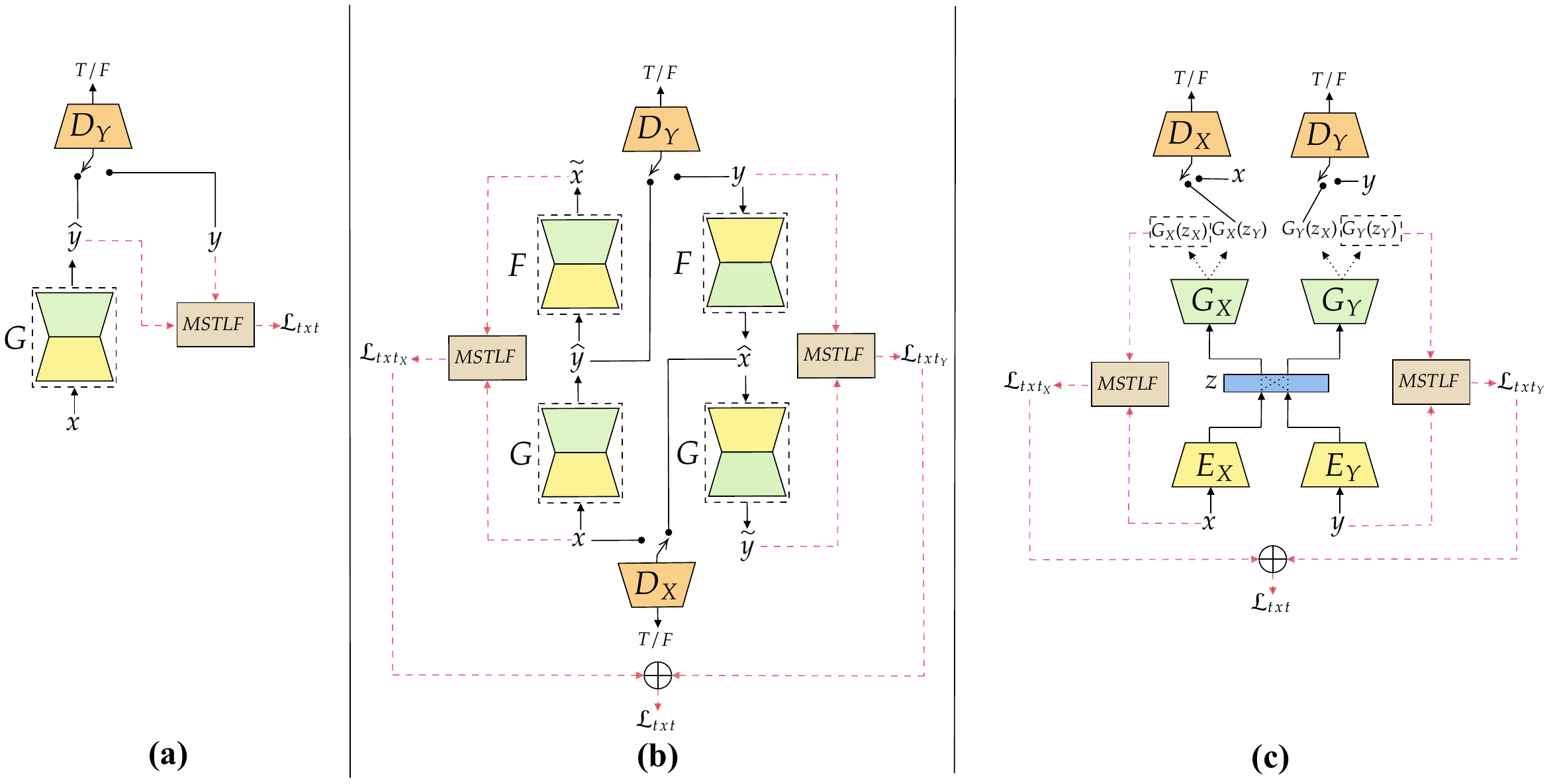}

 \caption{GAN architectures showing how MSTLF is applied. (a) Pix2Pix (b) CycleGAN (c) UNIT. The red dashed arrows denote the loss function computation. In UNIT, we use black dotted arrows to distinguish between different generation paths. $G$ and $F$ denotes generator networks, $D_{X}$ and $D_{Y}$ indicates discriminator networks,  whilst $E_{X}$ and $E_{Y}$ are encoders and $G_{X}$ and $G_{Y}$ are decoders.}
 \label{fig:backbones}
\end{figure*}

The last four loss functions shown in \tablename~\ref{tab:losses} are $\mathcal{L}_{txt}^{max}$, $\mathcal{L}_{txt}^{avg}$,$\mathcal{L}_{txt}^{Frob}$,$\mathcal{L}_{txt}^{att}$
which differ in the type of aggregation used to combine texture descriptors.
As mentioned in section \ref{sec:MSTLF}, we have four approaches in the AM: three are static (maximum, average and Frobenius) and they are denoted as $\mathcal{L}_{txt}^{max}$, $\mathcal{L}_{txt}^{avg}$, $\mathcal{L}_{txt}^{Frob}$, respectively.
The fourth, named $\mathcal{L}_{txt}^{att}$, is dynamic and it makes use of an attention mechanism. Before concluding this subsection, we detail how we apply MSTLF for each of the three GANs, graphically presented in Figure~\ref{fig:backbones}.

Without loss of generality, let $\mathcal{L}_{txt}$ be a generic MSTLF. 
In Pix2Pix, we apply MSTLF between a denoised image $\bm{\hat{y}}=G(\bm{x},\bm{z})$ and its reference ground truth $y$, i.e., $\mathcal{L}_{txt}(\bm{\hat{y}},\bm{y})$ (\figurename~\ref{fig:backbones} (a)). 
In CycleGAN, we apply MSTLF to enforce the cycle consistency constraint between an image and its reconstruction in both the domains, i.e., $\mathcal{L}_{txt}(\bm{\Tilde{x}},\bm{x})$, $\mathcal{L}_{txt}(\bm{\Tilde{y}},\bm{y})$ where $\bm{\Tilde{x}}=F(G(\bm{x}))$ and $\bm{\Tilde{y}}=G(F(\bm{y}))$ are reconstructed images in the source domain $X$ and the target domain $Y$, respectively (\figurename~\ref{fig:backbones} (b)). 
In UNIT, MSTLF is applied as, $\mathcal{L}_{txt}(\bm{\hat{x}},\bm{x})$, $\mathcal{L}_{txt}(\bm{\hat{y}},\bm{y})$ where $\bm{\hat{x}}=G_{X}(\bm{\hat{z}_{X}})$ and $\bm{\hat{y}}=G_{Y}(\bm{\hat{z}_{Y}})$ (\figurename~\ref{fig:backbones} (c)).
%-----------------------------------------------------------------------------------------------
\subsection{Materials}\label{sub:Materials}
We use three public datasets that contain paired and unpaired images.
The first is the Mayo Clinic LDCT-and-Projection Data~\citep{moen2021low} which includes thoracic and abdominal HDCT and corresponding LDCT data simulated by a quarter dose protocol. 
We took 17 patients; among them, 10 are those used in the Mayo Clinic for the 2016 NIH-AAPM-Mayo Clinic Low Dose CT Grand Challenge\footnote{\url{https://www.aapm.org/grandchallenge/lowdosect/}.} and 9 out of 10, corresponding to 5376 CT slices, are used for training as other works on LDCT denoising did ~\citep{li2020investigation, tan2022selective, ma2020low, li2021low}. 
The test set, denoted as {\em Dataset A} in the following, contains 2994 slices from 8 patients: 1 patient from the aforementioned 10, plus other 7 patients randomly selected from ~\citep{moen2021low} to make the evaluation more robust\footnote{The supplementary reports the list of patients' IDs used in this work from all three datasets.}.
The evaluation on the Mayo Clinic LDCT-and-Projection Data permits us to compute paired metrics reported in section~\ref{sub:metrics}, but it could be limited by the fact that LDCT images are simulated.
To overcome this limitation we consider two real LDCT datasets that, straightforwardly, do not contain HDCT images. 
They are the Lung Image Database Consortium and Image Database Resource Initiative (LIDC/IDRI)~\citep{armato2011lung} and the ELCAP Public Image Database~\citep{VisionIAGroup}, named {\em Dataset B} and {\em Dataset C} hereinafter.
From the former, we included 8 patients for a total of 1831 LDCT images, whilst from the latter we used the scans from 50 patients for a total of 12360 LDCT images.
%The former includes clinical thoracic CT scans acquired either with standard dose or low-dose protocols. 
%By looking at the metadata and specifically the tube current, which is the main parameter used for determining the dose of the emitted radiation, we selected 8 patients for a total of 1831 LDCT images with tube current $<80 mAs$.
%We also visually inspect the data to check the presence of noise within the images. 
%The latter contains only LDCT images of 50 patients for a total of 12360 LDCT images.
%-----------------------------------------------------------------------------------------------
%\subsection{Denoising Evaluation} \label{sub:metrics}
\subsection{Performance metrics} \label{sub:metrics}
In~\citep{10178770}, they brought out that paired and unpaired metrics generally lead to different but complementary results, that is, they measure image quality from different perspectives.
Here we used five image quality assessment metrics: Mean Square Error (MSE), Peak-Signal-to-Noise Ratio (PSNR), Structural Similarity Index (SSIM), Natural Image Quality Evaluator (NIQE), and Perception Based Image QUality Evaluator (PIQUE).
The first three metrics perform a paired evaluation, i.e., they require pairs of images to be computed, whilst the last two relax this constraint performing a no-reference evaluation ~\citep{10178770}.
Both PSNR and MSE serve as error-based metrics, appropriate for quantifying the degree of image distortion without necessarily implying visually favorable outcomes. 
While SSIM has been introduced as a more sophisticated metric compared to the aforementioned two, it is better suited for measuring distortion rather than perception ~\citep{blau2018perception,tronchin2021evaluating}.
%SSIM perceptual and semantic meaning has been questioned \cite{yang2018low, you2019ct} making it more suitable for quantifying distortion than perception \cite{blau2018perception}.
Oppositely, NIQE and PIQUE are blind image quality assessment scores that only make use of measurable deviations measuring the perception~\citep{mittal2012making,venkatanath2015blind}.
 
\subsubsection*{MSE}
MSE computes the absolute difference between intensities in pixels of the denoised and reference image:
\begin{equation}\label{eq:mse}
MSE(\bm{\hat{y}},\bm{y}) = \frac{1}{mn}\sum_{i=0}^{m-1}\sum_{j=0}^{n-1}\left[\bm{\hat{y}}(i,j) - \bm{y}(i,j)\right]^2
\end{equation}
where $(i,j)$ is the pixel coordinates, and $m$, $n$ are the numbers of columns and rows, respectively. It ranges from 0 to $+\infty$. A lower MSE indicates a better match between the denoised and the reference image.

\subsubsection*{PSNR}
PSNR compares the maximum intensity in the image ($MAX_{\bm{\hat{y}}}$) with the error between $\bm{\hat{y}}$ and $\bm{y}$ given by the MSE:
\begin{equation}\label{eq:psnr}
PSNR(\bm{\hat{y}},\bm{y}) = 10\times log_{10}\left(\frac{MAX_{\bm{\hat{y}}}^2}{MSE(\bm{\hat{y}},\bm{y})}\right).
\end{equation}
It ranges from 0 to $+\infty$ and it is expressed in decibels.
Higher PSNR values indicate better quality, while lower values may suggest more noticeable distortions or errors in the denoised image.

\subsubsection*{SSIM} SSIM computes the local similarity between two images as a function of luminance, contrast, and structure~\citep{wang2004image}:
\begin{equation}\label{eq:ssim}
SSIM(\bm{\hat{y}},\bm{y}) = \frac{(2\mu_{\bm{\hat{y}}}\mu_{\bm{y}} + C_{1})(2\sigma_{ \bm{\hat{y}}\bm{y}} + C_{2})}{(\mu_{\bm{\hat{y}}}^2 + \mu_{\bm{y}}^2 +C_{1})(\sigma_{\bm{\hat{y}}}^2 + \sigma_{\bm{y}}^2 + C_2)}
\end{equation}
where $\mu_{\bm{\hat{y}}}$, $\mu_{\bm{y}}$ are mean intensities of pixels in $\bm{\hat{y}}$ and $\bm{y}$, respectively; similarly $\sigma_{\bm{\hat{y}}}^2$ and $\sigma_{\bm{y}}^2$ are the variances, $\sigma_{\bm{\hat{y}}\bm{y}}$ is the covariance whilst $C_{1}$ and $C_{2}$ are constant values to avoid numerical instabilities. It lies in $ [0,1] $ with values closer to 1 indicating better image quality.

\subsubsection*{NIQE} 
NIQE uses measurable deviations from statistical regularities observed in a corpus of natural images to quantify image perception, fitting a multivariate Gaussian model ~\citep{mittal2012making}:
\begin{equation}
\begin{split}
NIQE(\nu_{1}, \nu_{\bm{\hat{y}}}, \Sigma_{1}, \Sigma_{\bm{\hat{y}}})\\ =
\left(\sqrt{(\nu_{1} - \nu_{\bm{\hat{y}}})^{T} \left(\frac{\Sigma_{1}+\Sigma_{\bm{\hat{y}}}}{2}\right)^{-1}(\nu_{1}-\nu_{\bm{\hat{y}}})}\right)\cdot100
\end{split}
\end{equation}
where $\nu_{1}$, $\Sigma_{1}$, $\nu_{\bm{\hat{y}}}$, $\Sigma_{\bm{\hat{y}}}$ are the mean vectors and covariance matrices of the natural multivariate Gaussian model and the denoised image's multivariate Gaussian model, respectively. 
It lies in $[0, 100]$, with values closer to $0$ indicating better image quality.

\subsubsection*{PIQUE} 
PIQUE measures perception in a denoised image exploiting local characteristics extracted from $16\times16$ non-overlapping blocks~\citep{venkatanath2015blind}:
\begin{equation}
\begin{split}
PIQUE(\bm{\hat{y}}) =\left(\frac{\left( \sum_{k=1}^{N_{SA}}D_{sk}\right)+C_{1}}{N_{SA}+C_{1}}\right)\cdot100
\end{split}
\end{equation}
where $D_{sk}$ captures the distortion of each image block, $N_{SA}$ is the number of distorted blocks and $C_{1}$ is a constant to prevent numerical instability. 
It lies in $[0, 100]$, with values closer to $0$ indicating better image quality.

%-----------------------------------------------------------------------------------------------
\subsection{Implementation Details} \label{subsec:implementationDetails}
We preprocess all the images as follows: first, we convert all the raw DICOM files to Hounsfield Unit (HU), second, we select a display window centered in -500 HU with a width of 1400 HU, emphasizing lung tissue, third, we normalize all the images in the range [-1, 1] and resized to $256 \times 256$.
The training dataset is paired, as mentioned in section \ref{sub:Materials}: hence, we trained the Pix2Pix in a paired manner, while we trained CycleGAN and UNIT in an unpaired manner by scrambling all the training images to avoid paired correspondence between HDCT and LDCT images for each training batch.
We trained all the models for 50 epochs, with a batch size of 16. All the networks are initialized with normal initialization and optimized by Adam optimizer~\citep{kingma2014adam} with default learning rates of $2\cdot10^{-4}$ for Pix2Pix and CycleGAN and $10^{-4}$ for UNIT.
We set the weights of the baseline loss functions according to the original implementations, as described in the supplementary material. 
The weights of perceptual losses, edge loss, and SSIM loss were set to $\lambda_{VGG} = 0.1$, $\lambda_{AE-CT} = 10^{3}$, $\lambda_{EDGE} = 10$, $\lambda_{SSIM} = 1$, respectively. The weights of MSTLFs were set to $\lambda_{txt}^{max}=\lambda_{txt}^{avg}=\lambda_{txt}^{Frob}=10^{-3}$, $\lambda_{txt}^{att}=1$.
The latter refers to MSTLF-attention and it was set to 1, as the self-attention layer incorporates its normalization and does not require external scaling. Our MSTLF used a set of spatial offsets $D = \{1, 3, 5, 7\}$ and a set of angular offsets $\Theta = \{0^{\circ}, 45^{\circ}, 90^{\circ}, 135^{\circ}\}$.
%The weights of perceptual and texture loss functions were empirically set to $\lambda_{VGG} = 0.1$, $\lambda_{txt}^{max} = 10^{-3}$,$\lambda_{txt}^{max} = 10^{-3}$, respectively. 
 All the models are parallelized on 4 NVIDIA Tesla A100, allowing for faster training time. We repeat each experiment three times to make the results more robust, mitigating the intrinsic variability that characterizes GANs' training. 
 We assess the average statistical significance of each experiment through the Wilcoxon Signed rank Test with $p < 0.05$. %a p-value threshold set to $p < 0.05$ and \ggr by taking into account the average scores for each patient (refer to section 5 in the supplementary for statistical analysis on a per-patient basis). \bb
 
%-----------------------------------------------------------------------------------------------
\section{Results and Discussion} \label{sec:ResultsDiscussion}
This section performs an in-depth analysis to assess the effects of our MSTLF on LDCT denoising, providing both quantitative evaluations and visual examples.

\begin{table*}[!h]
\centering
\begin{adjustbox}{center}
\resizebox{12cm}{!}{
\begin{tabular}{l|ll|ccccc}
\noalign{\hrule height 1pt}
\multirow{3}{8em}{\textbf{Backbone}} & & \multirow{2}{8em}{\hspace{-0.55cm}\textbf{Experiment}} & \multicolumn{5}{|c}{\textbf{Dataset A}} \\ \cline{4-8}
& & & \textbf{PSNR $\uparrow$} & \textbf{MSE $\downarrow$} & \textbf{SSIM $\uparrow$} & \textbf{NIQE $\downarrow$} & \textbf{PIQUE $\downarrow$} \\
\cline{2-8}
&LDCT reference&  & 17.7322 & $2.25\cdot10^{-2}$ & 0.5832 & 17.0272 &23.5495 \\ \hline\hline
& \multirow{5}{4em}{Competitors}&\hspace{0.5cm} Baseline & 21.3344 & $10.50\cdot10^{-3}$ & 0.7335 & 6.5263& 8.6017 \\
\multirow{7}{4em}{Pix2Pix}&& \hspace{0.5cm} VGG-16 & 21.3481 & $10.41\cdot10^{-3}$ & 0.7293 & 6.5673 &8.7114 \\
&& \hspace{0.5cm} AE-CT & 21.3422 & $10.55\cdot10^{-3}$ & 0.7349 & 6.1640 & 8.9156 \\
&& \hspace{0.5cm} SSIM-L & 21.4011 & $10.37\cdot10^{-3}$ & $\textbf{0.7426}^{*}$ & 6.0690 & 8.2995 \\ 
&& \hspace{0.5cm} EDGE & 21.0620 & $10.63\cdot10^{-3}$ & 0.7196 & 6.4413 & 8.7461 \\ \cline{2-8}
&& \hspace{0.5cm} MSTLF-max & 21.3234 & $10.50\cdot10^{-3}$ & 0.7284 & 6.5024 &8.7950\\ 
&\multirow{1}{4em}{Our approach}&\hspace{0.5cm}  MSTLF-average & $\textbf{21.4497}^{*}$ & $\mathbf{{10.32}\cdot10^{-3*}}$ & 0.7379 & 6.2366 &8.5473 \\
&& \hspace{0.5cm} MSTLF-Frobenius & 21.3661 & $10.42\cdot10^{-3}$ & 0.7347 & 6.0670 &8.8096 \\ 
&& \hspace{0.5cm} MSTLF-attention & 20.7495 & $11.62\cdot10^{-3}$ & 0.7103 & $\textbf{5.1934}^{*}$&$\textbf{7.3575}^{*}$ \\ 
\hline\hline
& \multirow{5}{4em}{Competitors}&\hspace{0.5cm} Baseline & 20.8588 & $10.15\cdot10^{-3}$ & 0.6935 & 9.6203 &12.6893\\ 
\multirow{7}{4em}{CycleGAN}&&\hspace{0.5cm} VGG-16 & 20.7759 & $10.17\cdot10^{-3}$ & 0.6997 &10.1728&13.3256 \\ 
&&\hspace{0.5cm} AE-CT & 21.2143 & $9.69\cdot10^{-3}$ & 0.7092 & 8.0317 &8.7166 \\
&&\hspace{0.5cm} SSIM-L & 20.6699 & $10.50\cdot10^{-3}$ & 0.6919 & 10.2733 &13.0187 \\ 
&&\hspace{0.5cm} EDGE & 20.5538 & $10.84\cdot10^{-3}$ & 0.6928 & 10.7092&13.5882 \\ \cline{2-8}
&&\hspace{0.5cm} MSTLF-max & 20.9735 & $9.70\cdot10^{-3}$ & 0.7013 &9.2843 &11.9323\\ 
&\multirow{1}{4em}{Our approach}&\hspace{0.5cm} MSTLF-average & $\textbf{21.2529}^{*}$ & $\mathbf{{9.32}\cdot10^{-3*}}$ & $\textbf{0.7114}^{*}$ & 8.6642 & 10.9737 \\ 
&&\hspace{0.5cm} MSTLF-Frobenius &21.1883 & $9.54\cdot10^{-3}$ &0.7108 &8.5100 &11.1255\\ 
&&\hspace{0.5cm} MSTLF-attention & 20.6741 & $9.89\cdot10^{-3}$ & 0.7058
 &$\textbf{5.3209}^{*}$&$\textbf{5.4432}^{*}$\\ 
\hline\hline
& \multirow{5}{4em}{Competitors}&\hspace{0.5cm} Baseline & 21.4333 & $8.85\cdot10^{-3}$
 & 0.7203 &6.4581 & 8.3179 \\ 
\multirow{7}{4em}{UNIT}&&\hspace{0.5cm} VGG-16 & 21.4207 &$8.54\cdot10^{-3}$ & 0.7296 & 6.5030 & 7.7908 \\
&&\hspace{0.5cm} AE-CT & 22.0836 & $7.68\cdot10^{-3}$ & 0.7450 & 7.2524 & 7.6388 \\
&&\hspace{0.5cm} SSIM-L & 22.1704 & $7.64\cdot10^{-3}$ & 0.7465 & 7.4184 & 7.5623 \\ 
&&\hspace{0.5cm} EDGE & 21.9671 & $7.83\cdot10^{-3}$ & 0.7434 & 6.6897 & $\textbf{7.2272}^{*}$ \\ \cline{2-8}
&&\hspace{0.5cm} MSTLF-max & $\textbf{22.2671}^{*}$ & $\mathbf{7.32\cdot10^{-3*}}$ & $\textbf{0.7480}^{*}$ &6.8441&7.5657 \\ 
&\multirow{1}{4em}{Our approach}&\hspace{0.5cm} MSTLF-average & 22.0836 & $7.70\cdot10^{-3}$ & 0.7429 &6.7520 & 7.8877 \\ 
&&\hspace{0.5cm} MSTLF-Frobenius & 22.1601 & $7.41\cdot10^{-3}$ & 0.7458 &7.1659 &7.7813 \\ 
&&\hspace{0.5cm} MSTLF-attention & 22.1270 & $7.39\cdot10^{-3}$ &0.7426 &$\textbf{6.0238}^{*}$ &7.4101 \\ 
\noalign{\hrule height 1pt}
\end{tabular}
}
\end{adjustbox}
\caption{Quantitative comparison of different loss function configurations on the simulated test set. The numerical values are averages over three repetitions of each experiment.
Results highlighted in bold with one asterisk $(*)$, indicate the best performance for each metric and architecture with statistical significance $(p<0.05)$.
} 
%The best results are marked in bold and their statistical significance is assessed with Wilcoxon signed rank test ($p < 0.05$).}
\label{tab:dataset_A}
\end{table*}

\begin{table*}[!h]
\begin{center}
\begin{adjustbox}{center}
\centering
\resizebox{12cm}{!}{
\begin{tabular}{l|ll|cc|cc}
\noalign{\hrule height 1pt}

\multirow{3}{8em}{\textbf{Backbone}} & & \multirow{2}{8em}{\hspace{-0.55cm}\textbf{Experiment}} & \multicolumn{2}{ c|}{\textbf{Dataset B}} & \multicolumn{2}{ c}{\textbf{Dataset C}}\\ \cline{4-7}
& & & \textbf{NIQE $\downarrow$} & \textbf{PIQUE $\downarrow$}& \textbf{NIQE $\downarrow$} & \textbf{PIQUE $\downarrow$} \\
\cline{2-7}
&LDCT reference&  & 13.6189 &20.3687 & 11.8081 &15.1818\\ \hline\hline
& \multirow{5}{4em}{Competitors}&\hspace{0.5cm} Baseline & 5.5597 & 8.9952 & 5.4423 & 6.9620 \\ 
\multirow{7}{4em}{Pix2Pix}&& \hspace{0.5cm} VGG-16 & 5.5726 & 8.8453 & 5.4671 & 6.8522 \\
&& \hspace{0.5cm} AE-CT & 5.1338 & 8.4187 & 5.0586 & 6.4835 \\
&& \hspace{0.5cm} SSIM-L & 5.1439 & 8.3272 & 5.1091 & $\textbf{6.2623}^{*}$ \\ 
&& \hspace{0.5cm} EDGE & 5.5052 & 9.5456 & 5.6078 & 7.4611 \\ \cline{2-7}
&& \hspace{0.5cm} MSTLF-max & 5.5234 & 9.0999 & 5.4585 & 7.061\\ 
&\multirow{1}{4em}{Our approach}&\hspace{0.5cm} MSTLF-average & 5.2658 & 8.7173 & 5.2390 &6.6374 \\
&& \hspace{0.5cm} MSTLF-Frobenius & 5.1832 & 8.5485 & 5.1944 & 6.7230 \\ 
&& \hspace{0.5cm} MSTLF-attention & $\textbf{4.8728}^{*}$ & $\textbf{7.3443}^{*}$ & $\textbf{4.8634}^{*}$ & $\textbf{6.2841}^{*}$ \\ 
\hline\hline
& \multirow{5}{4em}{Competitors}&\hspace{0.5cm} Baseline & 8.6468 & 12.6395 &8.3811 & 10.5286\\
\multirow{7}{4em}{CycleGAN}&&\hspace{0.5cm} VGG-16 & 8.3012 & 10.9972 & 8.1010 &9.0185 \\ 
&&\hspace{0.5cm} AE-CT & 7.1581 & 7.3931 & 6.9573 &6.5963 \\
&&\hspace{0.5cm} SSIM-L & 8.9653 & 12.2665 & 8.7280 & 10.2850 \\ 
&&\hspace{0.5cm} EDGE & 8.9325 & 12.4375 & 8.6059 & 10.3436 \\ \cline{2-7}
&&\hspace{0.5cm} MSTLF-max & 8.3655 & 11.7319 & 8.1205 &9.9322\\ 
&\multirow{1}{4em}{Our approach}&\hspace{0.5cm} MSTLF-average & 7.9853 & 11.6693 & 7.8831 & 9.9328 \\ 
&&\hspace{0.5cm} MSTLF-Frobenius & 7.5627 &10.8876 &7.4276 &9.0732 \\ 
&&\hspace{0.5cm} MSTLF-attention & $\textbf{5.1713}^{*}$ & $\textbf{6.2002}^{*}$ & $\textbf{5.1967}^{*}$ & $\textbf{5.7071}^{*}$\\ 
\hline\hline
& \multirow{5}{4em}{Competitors}&\hspace{0.5cm} Baseline & 5.9253 & 9.0455 & 5.9474 & 7.4220 \\ 
\multirow{7}{4em}{UNIT}&&\hspace{0.5cm} VGG-16 & 5.9518 & 9.0618 & 5.9586 & 7.4170\\
&&\hspace{0.5cm} AE-CT & 6.7394 & 8.8963 & 6.6581 & 7.1964 \\
&&\hspace{0.5cm} SSIM-L & 7.0960 & 8.5735 & 7.0274 & 6.9753 \\ 
&&\hspace{0.5cm} EDGE & 6.4342 & $\textbf{7.9593}^{*}$ & 6.3607 & $\textbf{6.5470}^{*}$ \\ \cline{2-7}
&&\hspace{0.5cm} MSTLF-max & 6.5491 & 8.4952 & 6.5070 & 6.8747 \\ 
&\multirow{1}{4em}{Our approach}&\hspace{0.5cm} MSTLF-average & 6.4214 & 9.1836 & 6.4317 & 7.4392 \\ 
&&\hspace{0.5cm} MSTLF-Frobenius & 6.6947 & 8.9574 & 6.6231 & 7.2449 \\ 
&&\hspace{0.5cm} MSTLF-attention & $\textbf{5.5228}^{*}$ & 8.5232
 & $\textbf{5.5433}^{*}$ & 7.0860 \\ 
\noalign{\hrule height 1pt}
\end{tabular}
}
\end{adjustbox}
\end{center}
\caption{ 
Quantitative comparison of different loss function configurations on the real test sets.  
The numerical values are averages over three repetitions of each experiment.
Results marked in bold with one asterisk $(*)$, indicate the best performance per metric and architecture with statistical significance $(p<0.05)$. Two results marked in bold with one asterisk $(*)$ within the same backbone section denote that they both satisfy $(p<0.05)$ wrt the other experiments while  $p\geq 0.05$ among themselves.
%Bold text and a single asterisk highlight the best results in a section with $p\geq 0.05$ among themselves, while satisfying $p < 0.05$ wrt to the other experiments.
}
\label{tab:dataset_B_C}
\end{table*}

%--------------------------------------------------------------------------
%\subsection{Effects of MSTLF on LDCT Denoising}
%We adopt five widely used IQA metrics for quantitative evaluation, including paired metrics (PSNR, MSE, and SSIM) and unpaired metrics (NIQE and PIQE) that we describe in Section \ref{sub:metrics}. 
\tablename~\ref{tab:dataset_A} and \tablename~\ref{tab:dataset_B_C} present the results of different experiments on simulated LDCT data ({\em Dataset A}), and real LDCT data ({\em Dataset B} and {\em C}), respectively. 
The two tables share the same structure: as a reference, the first row shows the values of the metrics computed on the LDCT images, which are used hereinafter to evaluate the relative improvement in image quality.
The following rows are grouped in sections given the GAN backbone. 
For each section, there are nine experiments corresponding to the loss functions presented in section \ref{sec:Competitors} and divided into five competitors and four implementations of the proposed approach, differing in the type of aggregation employed.
%For each section, there are nine experiments corresponding to the loss functions presented in section \ref{sec:Competitors} and summarized in \tablename~\ref{tab:losses}. 
In each section, we employ bold text and two asterisks to emphasize the best results per metric that exhibit statistically significant differences from the others, satisfying $p < 0.05$.
%Additionally, we use bold text and a single asterisk to highlight the best results in a section that does not show statistically significant differences among themselves, while still satisfying $p < 0.05$ compared to the other experiments.

%All GAN-based models are characterized by inherent variability due to adversarial training which might compromise the reliability of the results. To address this issue, we show the results as the average of three repetitions for each experiment, assessing
%statistical significance through the Wilcoxon Signed Rank Test with p-value threshold set to $p < 0.05$. 
Both \tablename~\ref{tab:dataset_A} and \ref{tab:dataset_B_C} show that the proposed MSTLF approach performs better than the other loss functions on all the architectures in most of the cases: indeed, MSTLF has the largest number of bold values denoting the best results per metric and per architecture.
For each GAN backbone, the Wilcoxon Signed rank Test shows that these best results are always statistically different (except for one case in \tablename~\ref{tab:dataset_B_C}) from the others.
This finding is also confirmed when performing  the same test  on a per-patient basis, as shown in section 2 of the supplementary.

In Pix2Pix, MSTLF-average outperforms the other loss function configurations in terms of PSNR and MSE, while SSIM-L reaches the highest SSIM score, which is expected since SSIM-L is meant to maximize the SSIM. 
MSTLF-attention achieves the best result in NIQE and PIQUE on {\em Dataset A}, {\em B}, and {\em C}, 
proving to be effective on both simulated LDCT data and real LDCT data.

Turning our attention to the results attained using the CycleGAN, the results are consistent with those of Pix2Pix: MSTLF-average outperforms the other loss function configurations in all paired metrics while MSTLF-attention reaches the best performance in terms of unpaired metrics (NIQE and PIQUE) on all the datasets. 

\begin{figure}[h]
\centering
\includegraphics[width=9cm]{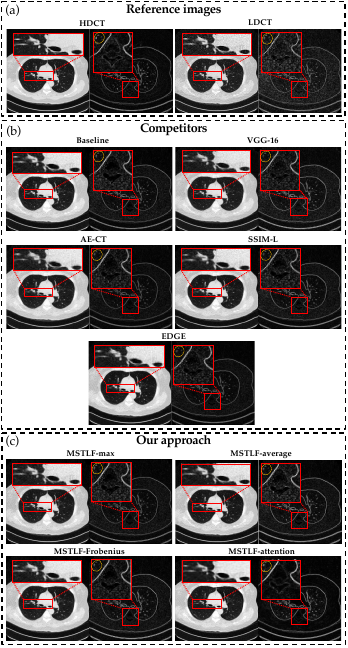}
\caption{
Visual comparison of denoised CT slices from the Mayo simulated data ({\em Dataset A}). 
We show all loss function configurations for CycleGAN (see supplementary for the remaining figures relative to CycleGAN, Pix2Pix and UNIT). 
The images are organized into ``Reference Images'' (panel(a)), ``Competitors'' (panel (b)) and ``Our Approach'' (panel (c)). For each configuration, results are displayed in the image domain (left image, with a display window of [-1200, 200] HU) and the gradient domain (right image).
Zoomed ROIs highlight key regions demonstrating noise reduction effectiveness.
%Visual comparison of denoised CT slices from the Mayo simulated data ({\em Dataset A}). 
%We show all loss function configurations for CycleGAN (see supplementary for the remaining figures relative to CycleGAN, Pix2Pix and UNIT). 
%The images are shown both in the image domain (panel (a) using a  display window in [-1200, 200] HU), and the gradient domain (panel (b)). 
%The zoomed ROI in panel (c) highlights denoising in the gradient domain.
}
\label{fig:cycleGANimages}
\end{figure}

In UNIT, MSTLF-max is the best configuration focusing on paired metrics, while MSTLF-attention stands out in NIQE, as in Pix2Pix and CycleGAN. The best PIQUE instead is achieved by EDGE which, however, does not excel in any other metrics or architectures.
On the contrary, MSTLF shows consistency of results on most metrics regardless of the architecture,
showcasing a higher degree of agnosticism compared to other loss functions.
We further observe that configurations that perform a static aggregation, i.e., MSTLF-max, MSTLF-average, MSTLF-Frobenius, favor paired metrics, whilst MSTLF-attention, performing dynamic aggregation, favors unpaired metrics.
\figurename~\ref{fig:cycleGANimages} shows an example of denoised images on \emph{Dataset A} using the CycleGAN backbone, where   
 panel (a)  reports the HDCT and LDCT images.
 Panel (b) then presents the results of the five competitors, 
whilst panel (c) shows the results attained by the four configurations of our approach.
For each configuration (HDCT, LDCT, VGG-16 and so on), the figure offers  images both in the image domain (left) and the gradient domain (right) to highlight noise characteristics. 
We also include zoomed-in Region of Interests (ROIs) marked with red boxes, capturing areas inside and outside the lung regions to highlight various tissues and anatomical structures.
It is worth noting that MSTLF-attention (bottom-right image) produces the image with the lowest gradient value, indicating the most effective noise reduction while preserving the lung boundary details.
Such a result confirms the effectiveness of dynamically aggregating texture information during training.
The other three configurations, i.e., MSTLF-max, MSTLF-average, and MSTLF-Frobenius, while not outperforming all competitors in noise reduction, excel in retaining edge textures.
In contrast, the competitors, particularly AE-CT, SSIM-L, and EDGE, struggle to preserve edge details, leading to blurred lung boundaries and suboptimal noise reduction.
This is particularly evident in the yellow circle, where fine-grained details are well-maintained across all MSTLF implementations, further confirming the robustness of our approach.
Section 4 in the supplementary material completes the visual examples presenting the results for the other two datasets and GAN backbones.
%In panel (a) we report the HDCT and LDCT images followed by synthetic images provided by the nine experiments, whilst panel (b) shows the gradient of these images, to highlight noise in the image. 
%For the sake of visualization, panel (c) shows the zoomed Regions Of Interest (ROIs) identified by the red boxes in the gradient images, which contain areas inside and outside the lung regions, thus including different tissues and anatomical structures.
%In the first and third rows of the figure, we report the HDCT and LDCT images followed by synthetic images provided by the nine experiments.
%The second and fourth rows show the gradient of these images, a visualization stratagem that highlights the noise in the image.
%or the sake of visualization, the fifth and sixth rows show the zoomed regions identified by the red boxes in the gradient images, which contain areas inside and outside the lung regions thus including different tissues and anatomical structures. 
%It is worth noting that MSTLF-attention produced the image with the lowest gradient that, in turn, corresponds to the lowest level of noise while preserving the lung boundaries (MSTLF-attention in \figurename~\ref{fig:cycleGANimages} (c)). 

\FloatBarrier
\subsection{The Perception-Distortion Trade-Off}
\begin{figure*}[h]
\begin{center}
\begin{adjustbox}{center}
\centering
\includegraphics[width=9cm]{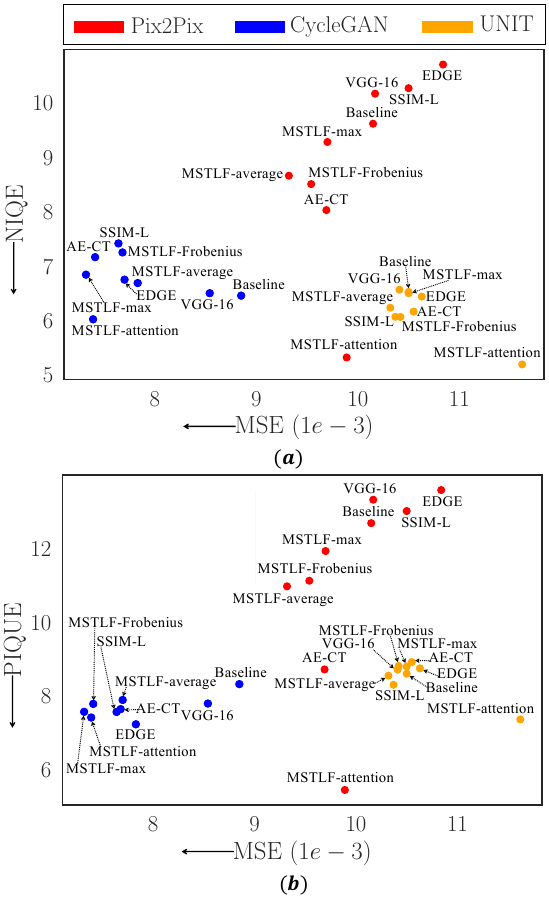}
\label{fig:img1}
% \end{subfigure}
% \hfill
% \begin{subfigure}[b]{0.45\textwidth}
% \centering
% \includegraphics[width=\textwidth]{PIQE-MSE.pdf}
% \caption{PIQUE vs MSE}
% \label{fig:img2}
% \end{subfigure}
\end{adjustbox}
\end{center}
 \caption{Perception-distortion evaluation of denoising algorithms. We plot the 9 experiments detailed in ~\tablename~\ref{sec:Competitors} on the perception distortion plane. 
 The color legend groups the experiments based on the GAN architecture. Distortion is measured by MSE in both panels, whilst perception is measured by NIQE and PIQUE in panel (a) and panel (b), respectively.}
 \label{fig:perc-dist}
\end{figure*}

~\citep{blau2018perception} demonstrated that there exists a trade-off between perception and distortion when dealing with image restoration algorithms, that is, a low average distortion does not necessarily imply a good perceptual quality, and achieving both is an open challenge. 
From this consideration, we now look at our results from a perception-distortion perspective; as reported in section~\ref{sub:metrics}, perception is measured by NIQE and PIQUE, whereas distortion is measured by the MSE. 
Accordingly, we represent each experiment by a point in the two perception-distortion planes (panels (a) and (b) of \figurename~\ref{fig:perc-dist}).
Then, we assess the trade-off between perception and distortion by ranking the experiments based on their proximity to the origin, as the point $(0,0)$ is the ideal one (\tablename~\ref{tab:ranking}). 
We find that MSTLF-attention is the one that reaches the best trade-off between perception and distortion, demonstrating consistent performance across different architectures. 
This outcome further supports the agnostic nature of our approach with respect to the GAN architecture.

\begin{table*}[!h]
\begin{center}
\begin{adjustbox}{center}
\centering
\resizebox{12.5cm}{!}{
\begin{tabular}{lcccc@{ \vline }ccccc}
%\begin{tabular}{lcccc|cccc}
\noalign{\hrule height 1pt}
\multirow{2}{8em}{\textbf{Experiment}} & \multicolumn{4}{c@{ \vline }}{\textbf{NIQE vs MSE ranking}} & \multicolumn{4}{c}{\textbf{PIQUE vs MSE ranking}}\\ \cline{2-9}

& \textbf{Pix2Pix} & \textbf{CycleGAN} & \textbf{UNIT} & \textbf{Global} & \textbf{ Pix2Pix} & \textbf{CycleGAN} & \textbf{UNIT} & \textbf{Global}\\
\noalign{\hrule height 1pt}
Baseline & 8 & 6 & 2 & 45 & 4 & 6 & 9 & 50 \\
VGG-16 & 9 & 7 & 3 & 49 & 5 & 8 & 7 & 50 \\
AE-CT & 4 & 2 & 8 & 26 & 9 & 2 & 5 & 36 \\
SSIM-L & 3 & 8 & 9 & 31 & 2 & 7 & 3 & 36 \\ 
EDGE & 6 & 9 & 4 & 35 & 6 & 9 & \textbf{1} & 44 \\ 
MSTLF-max & 7 & 5 & 6 & 49 & 7 & 5 & 4 & 47\\ 
MSTLF-average & 5 & 4 & 5 & 44 & 3 & 3 & 8 & 44 \\
MSTLF-Frobenius & 2 & 3 & 7 & 42 & 8 & 4 & 6 & 49\\ 
MSTLF-attention & \textbf{1} & \textbf{1} & \textbf{1} & \textbf{6} & \textbf{1} & \textbf{1} & 1 & \textbf{8}\\ 
\noalign{\hrule height 1pt}
\end{tabular}
}
\end{adjustbox}
\end{center}

\caption{Distance-based ranking in the perception-distortion plane. A lower rank denotes a better trade-off between perception and distortion.}
\label{tab:ranking}
\end{table*}
\FloatBarrier
\subsection{Template Matching Analysis}
To gain more insight into the denoising performance, we perform an analysis that quantifies the LDCT noise pattern still present in the corresponding denoised version. 
To this aim, let us remember that $\bm{x}$ is a LDCT image resized to $256 \times 256$ (see section~\ref{subsec:implementationDetails}), now belonging to a set of $d$ test images.
Furthermore, $\bm{T} \in \bm{x}$ denotes a template of size $t \times t$ (with $t=32$) extracted from $\bm{x}$ at a given location. 
To quantify the aforementioned noise pattern %still present in the denoised version of the input image 
we proceed as follows. 
First, we compute the maximum template matching between $\bm{T}$ and the denoised image $\bm{\hat{y}}$ given by 
\begin{equation}
 m = \text{max}(\bm{M}(x, y)) \, \, \text{with} \, x,y \in [0,255]
\end{equation}
where $\bm{M}(x, y)$ is the template matching at location $(x,y)$ calculated using zero-padding as
\begin{equation}
\begin{split}
\bm{M}(x, y) = \frac{\sum_{x'y'}\bm{\bm{T}}(x',y')\cdot \bm{\hat{y}}(x+x',y+y')}{\sqrt{\sum_{x'y'}\bm{T}(x',y')^{2}\cdot \sum_{x'y'} \bm{\hat{y}}(x+x',y+y')^{2}}}
\end{split}
\end{equation}
noticing that $\bm{\hat{y}}$ stands for the denoised image, and $x'$, $y'$ represent the discrete unitary displacement, ranging in $[0, t-1]$, with respect to the current position $(x,y)$.
Hence, the numerator computes the cross-correlation between $\bm{T}$ and $\bm{\hat{y}}$, whilst the denominator is a normalization term.

Second, in order not to limit the analysis to a single image region in the original LDCT, we extract $r=9$ equispaced templates obtaining $r$ matching scores $m$.
Given $d$ LDCT images, we collect $r\cdot d$ matching scores that we use to estimate the underlying matching distribution through kernel density estimation:
%and we concatenate the matching scores in a single vector:
%\begin{equation}
% \bm{\mathcal{C}} = [m_{1}, m_{2},...,m_{r\cdot d}]
%\end{equation}
%where $d$ indicates the number of test images.\\
%Then, we 
\begin{equation}
\hat{f}(m) = \frac{1}{(r\cdot d)\cdot h}\sum_{i=1}^{r\cdot d} K\left( \frac{ m-m_{i}}{h} \right) 
\end{equation}
where $\hat{f}(m)$ is the estimated probability density function (PDF), %$r$ is the number of templates , $d$ is the size of the test set, $m$ is a generic matching score, 
$K$ is a Gaussian kernel function and $h$ is a non-negative smoothing parameter that can be estimated using Scott's rule (\cite{scott2015multivariate}).

\begin{figure*}[h]
\begin{center}
\begin{adjustbox}{center}
\centering
\includegraphics[width=12.5cm]{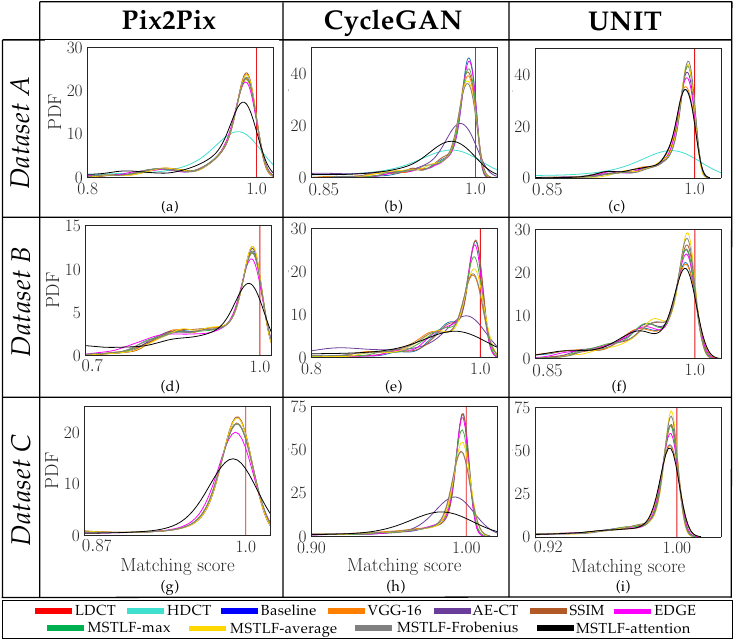}
% \end{subfigure}
% \hfill
% \begin{subfigure}[b]{0.45\textwidth}
% \centering
% \includegraphics[width=\textwidth]{images/PIQE-MSE.pdf}
% \caption{PIQUE vs MSE}
% \label{fig:img2}
% \end{subfigure}
\end{adjustbox}
\end{center}
 \caption{Template matching analysis. Each panel illustrates the matching distributions between the noisy templates extracted from the LDCT images and the denoised images.}
 \label{fig:tm}
\end{figure*}

\figurename~\ref{fig:tm} presents the results and is organized as follows: each row corresponds to a dataset from \emph{A} to \emph{C}, and it contains three panels, one for each GAN. In each panel, the matching distribution computed over the LDCT images is a delta function (in red) since all the templates come from the same LDCT images and, so, the matching scores are all equal to 1. In the first row we use a paired dataset and we, therefore, report in light blue the matching distribution between the LDCT templates and the HDCT images, which provides us with a lower boundary reference to assess any other matching distribution computed between the LDCT templates and the denoised images.
The visual inspection of the 9 plots confirms that MSTLF-attention is the best configuration. Indeed, looking at ~\figurename~\ref{fig:tm} (a)-(b)-(c), MSTLF-attention turns out to be the distribution that tends more toward the high-dose reference. Even in ~\figurename~\ref{fig:tm} (d)-(e)-(f)-(g)-(h)-(i), that do not report the high-dose reference, we notice that MSTLF-attention deviates more than other approaches from the low-dose reference.
The plots in ~\figurename~\ref{fig:tm} (b)-(e)-(h) also reveal that AE-CT has the closest behavior to MSTLF-attention, thus supporting the effectiveness and superiority of domain-specific deep features over ImageNet features. 
Indeed, the widely adopted VGG-16 perceptual loss does not excel in any of our analyses. Nonetheless, AE-CT is second in position to MSTLF-attention only in CycleGAN and if we look at the previous analysis it excels only in terms of NIQE (Table \ref{tab:dataset_A} and \ref{tab:dataset_B_C}) without reaching a satisfying perception-distortion trade-off (\figurename~\ref{fig:perc-dist} and Table \ref{tab:ranking}). 
In ~\figurename~\ref{fig:tm} (c) we observe that all configurations, including MSTLF-attention, remain distant from the high-dose boundary. We interpret this result as the tendency for UNIT to favor distortion rather than perception. Indeed, among the architectures, UNIT is the one that reaches the best PSNR, MSE, and SSIM (see Table \ref{tab:dataset_A}). 

%--------------------------------------------------------------------------
\subsection{Computational Analysis}
Table \ref{tab:times} shows the computational analysis, where the computation time is given by the average time over one training epoch needed to compute the loss function and to back-propagate its gradient. In general, both CycleGAN and UNIT are more 
computationally demanding ($\sim70\%$ more than Pix2Pix): this is expected since these two GANs have larger architectures than Pix2Pix, and also because they compute the loss term twice when performing the bidirectional translation between LD and HD domains.
Both VGG-16 and AE-CT cause an increase in computational burden compared to the baseline (up to $+36\%$) since they use a VGG-16 network and an auto-encoder as feature extractors, respectively. Although these networks are frozen, the back-propagation is always computed, hence affecting the overall computation time. On the contrary, SSIM-L and EDGE have the same computation time as the baseline since they introduce two loss terms that rely on mathematical operations applied to the images without additional cost for back-propagating the gradient. Turning our attention to our MSTLFs, these cause a significant increase in computation time (up to $+65\%$ with respect to the baseline) due to the complexity of the operations that extract a multi-scale texture representation exploiting the soft GLCMs. Indeed, ensuring compatibility with the gradient-based optimization frameworks leads to an implementation that is more computationally onerous than the standard GLCM (section \ref{sec: differentability}). Nonetheless, we notice that the type of aggregation used does not significantly affect the computational burden. In this regard, static aggregation, which is defined by mathematical operators, has the same impact as dynamic aggregation defined by self-attention. Such observation combined with the good performance reached by MSTLF-attention makes dynamic aggregation 
a promising approach.
\begin{table}[h]
\centering
\begin{tabular}{|c|c|c|c|c|}
\hline
%\multicolumn{2}{c|}{OBJECTIVE FUNCTION}
\textbf{Experiment} & \textbf{Pix2Pix} & \textbf{CycleGAN} & \textbf{UNIT} \\
\hline
Baseline & 0.09 s & 0.42 s & 0.28 s \\
\hline
VGG-16 & 0.13 s& 0.49 s & 0.44 s\\
\hline
AE-CT & 0.11 s& 0.47 s & 0.41 s\\
\hline
SSIM & 0.09 s& 0.42 s & 0.28 s\\
\hline
EDGE & 0.09 s& 0.42 s & 0.28 s\\
\hline
MSTLF-max & 0.26 s & 0.89 s& 0.81 s\\
\hline
MSTLF-average & 0.26 s & 0.94 s & 0.82 s\\
\hline
MSTLF-Frobenius & 0.26 s & 0.95 s & 0.82 s\\
\hline
MSTLF-attention & 0.26 s& 1.01 s & 0.84 s\\
\hline
\end{tabular}
\caption{Computational analysis. The computation time is expressed as the average time required to compute the loss function and to back-propagate its gradient for each batch.}
\label{tab:times}
\end{table}

%--------------------------------------------------------------------------
%--------------------------------------------------------------------------
\section{Conclusions} \label{sec:Conclusions}
%In this paper, we present an approach, named MSTLF, for GAN-based denoising in CT imaging that embeds textural information, extracted at different spatial and angular scales. 
\begin{comment}
\cc 
In this paper we present an approach, named MSTLF, for GAN-based denoising in CT imaging whose core methodological novelty   twofold: (1) its multi-scale approach to texture analysis that embeds textural information, extracted at different spatial and angular scales,  (2) its advanced aggregation methods. 
For the former point, we also overcome the limitations posed by the non-differentiability of the GLCM by proposing a novel implementation based on soft assignment that makes it compatible with gradient-based optimization frameworks. 
For the latter point, we investigate two types of aggregation, one static and one dynamic. 
We find that the dynamic one, which employs an attention mechanism,  enables the model to dynamically weigh and integrate information from multiple scales, resulting in a more context-sensitive and powerful representation of texture across different scales — a technique absent from existing literature, to the best of our knowledge.
\end{comment}
In this paper, we introduce an approach named MSTLF, designed for GAN-based denoising in CT imaging. 
The core methodological contributions are twofold: (1) a multi-scale approach to texture analysis that integrates textural information extracted at various spatial and angular scales, and (2) advanced aggregation methods to enhance the representation.
For the first contribution, we address the challenge of the GLCM’s non-differentiability by proposing a novel soft-assignment-based implementation. 
This approach ensures compatibility with gradient-based optimization frameworks, enabling seamless integration into modern deep learning pipelines.
For the second contribution, we explore two aggregation strategies: static and dynamic. Notably, the dynamic aggregation employs an attention mechanism, allowing the model to adaptively weight and integrate information across scales. 
This results in a context-sensitive and expressive representation of texture that, to the best of our knowledge, is absent in existing literature.
%We also overcome the limitations posed by the non-differentiability of the GLCM proposing a novel implementation based on soft assignment that makes it compatible with gradient-based optimization frameworks. 
%%The multiscale nature of our approach is fully exploited by the dynamic aggregation based on self-attention which not only allows MSTLF to achieve a good compromise in terms of distortion and perception but also makes it an architecture-independent module. With the surge of denoising frameworks, the availability of model-agnostic modules that can be interchangeably used across different architectures is an excellent resource that facilitates both the deployment and validation phases.
Our comprehensive analysis on three datasets and three GANs backbones brings out the effectiveness of MSTLF to enhance the quality of LDCT images. 
In particular, the configuration exploiting the attention layer to dynamically aggregate multi-scale texture data attains the best performance compared to other well-established loss functions. 
We also find two other interesting results.
First, it also successfully tackles the challenging trade-off between distortion and perception. 
%In this regard, the use of self-attention is crucial as it enables the dynamic aggregation of multiscale information while training.
Second, just as it tailors the aggregation to the specific needs of the optimization process, it also succeeds in adapting the GAN architecture used. %Indeed, all our analyses show consistent behaviors for MSTLF attention across the architectures corroborating the hypothesis that our approach is agnostic to the latter.
Future works are directed towards further investigation of the backbone-independent feature of MSTLF, exploring its integration with other denoising frameworks and generative model families in both medical and non-medical domains. 
Additionally, efforts will be directed toward optimizing its computational efficiency to make it suitable for real-time applications in resource-constraint environments.

%Such a feature makes MSTLF an architecture-independent module suitable for application and testing in a wide range of denoising frameworks in medical and non-medical applications. 
%In the future, we plan to thoroughly investigate the agnosticism of our approach by extending it to other tasks and architectures. Moreover, the promising results obtained with the dynamic aggregation motivate us to further study the use of self-attention in the context of hyperparameter optimization, an area where the development of effective end-to-end and adaptive techniques would be of high interest.

\section*{Acknowledgments} 
Resources are provided by the National Academic Infrastructure for Supercomputing in Sweden (NAISS) and the Swedish National Infrastructure for Computing (SNIC) at Alvis @ C3SE, partially funded by the Swedish Research Council through grant agreements no. 2022-06725 and no. 2018-05973.
We acknowledge financial support from: i) Cancerforskningsfonden Norrland project MP23-1122, ii) PNRR MUR
project PE0000013-FAIR, iii) PRIN 2022 MUR 20228MZFAA-AIDA (CUP C53D23003620008), iv) PRIN PNRR 2022 MUR P2022P3CXJ-PICTURE (CUP C53D23009280001), v) MISE FONDO PER LA CRESCITA SOSTENIBILE Bando Accordo Innovazione DM 24/5/2017 (CUP B89J23000580005).

%%Harvard
\bibliographystyle{model2-names.bst}\biboptions{authoryear}
\bibliography{refs}

\begin{thebibliography}{60}
\expandafter\ifx\csname natexlab\endcsname\relax\def\natexlab#1{#1}\fi
\providecommand{\url}[1]{\texttt{#1}}
\providecommand{\href}[2]{#2}
\providecommand{\path}[1]{#1}
\providecommand{\DOIprefix}{doi:}
\providecommand{\ArXivprefix}{arXiv:}
\providecommand{\URLprefix}{URL: }
\providecommand{\Pubmedprefix}{pmid:}
\providecommand{\doi}[1]{\href{http://dx.doi.org/#1}{\path{#1}}}
\providecommand{\Pubmed}[1]{\href{pmid:#1}{\path{#1}}}
\providecommand{\bibinfo}[2]{#2}
\ifx\xfnm\relax \def\xfnm[#1]{\unskip,\space#1}\fi
%Type = Article
\bibitem[{Aharon et~al.(2006)Aharon, Elad and Bruckstein}]{aharon2006k}
\bibinfo{author}{Aharon, M.}, \bibinfo{author}{Elad, M.}, \bibinfo{author}{Bruckstein, A.}, \bibinfo{year}{2006}.
\newblock \bibinfo{title}{{K-SVD: An algorithm for designing overcomplete dictionaries for sparse representation}}.
\newblock \bibinfo{journal}{IEEE Transactions on signal processing} \bibinfo{volume}{54}, \bibinfo{pages}{4311--4322}.
%Type = Article
\bibitem[{Armato~III(2011)}]{armato2011lung}
\bibinfo{author}{Armato~III, Samuel~G, M.}, \bibinfo{year}{2011}.
\newblock \bibinfo{title}{{The lung image database consortium (LIDC) and image database resource initiative (IDRI): a completed reference database of lung nodules on CT scans}}.
\newblock \bibinfo{journal}{Medical physics} \bibinfo{volume}{38}, \bibinfo{pages}{915--931}.
%Type = Inproceedings
\bibitem[{Bera and Biswas(2023)}]{bera2023self}
\bibinfo{author}{Bera, S.}, \bibinfo{author}{Biswas, P.K.}, \bibinfo{year}{2023}.
\newblock \bibinfo{title}{{Self Supervised Low Dose Computed Tomography Image Denoising Using Invertible Network Exploiting Inter Slice Congruence}}, in: \bibinfo{booktitle}{Proceedings of the IEEE/CVF Winter Conference on Applications of Computer Vision}, pp. \bibinfo{pages}{5614--5623}.
%Type = Article
\bibitem[{Bevelacqua(2010)}]{bevelacqua2010practical}
\bibinfo{author}{Bevelacqua, J.J.}, \bibinfo{year}{2010}.
\newblock \bibinfo{title}{{Practical and effective ALARA}}.
\newblock \bibinfo{journal}{Health physics} \bibinfo{volume}{98}, \bibinfo{pages}{S39--S47}.
%Type = Inproceedings
\bibitem[{Blau and Michaeli(2018)}]{blau2018perception}
\bibinfo{author}{Blau, Y.}, \bibinfo{author}{Michaeli, T.}, \bibinfo{year}{2018}.
\newblock \bibinfo{title}{{The perception-distortion tradeoff}}, in: \bibinfo{booktitle}{Proceedings of the IEEE conference on computer vision and pattern recognition}, pp. \bibinfo{pages}{6228--6237}.
%Type = Article
\bibitem[{Brenner and Hall(2007)}]{brenner2007computed}
\bibinfo{author}{Brenner, D.J.}, \bibinfo{author}{Hall, E.J.}, \bibinfo{year}{2007}.
\newblock \bibinfo{title}{{Computed tomography—an increasing source of radiation exposure}}.
\newblock \bibinfo{journal}{New England journal of medicine} \bibinfo{volume}{357}, \bibinfo{pages}{2277--2284}.
%Type = Book
\bibitem[{Burgos and Svoboda(2022)}]{burgos2022biomedical}
\bibinfo{author}{Burgos, N.}, \bibinfo{author}{Svoboda, D.}, \bibinfo{year}{2022}.
\newblock \bibinfo{title}{{Biomedical Image Synthesis and Simulation: Methods and Applications}}.
\newblock \bibinfo{publisher}{Academic Press}.
%Type = Inproceedings
\bibitem[{Charbonnier et~al.(1994)Charbonnier, Blanc-Feraud, Aubert and Barlaud}]{charbonnier1994two}
\bibinfo{author}{Charbonnier, P.}, \bibinfo{author}{Blanc-Feraud, L.}, \bibinfo{author}{Aubert, G.}, \bibinfo{author}{Barlaud, M.}, \bibinfo{year}{1994}.
\newblock \bibinfo{title}{{Two deterministic half-quadratic regularization algorithms for computed imaging}}, in: \bibinfo{booktitle}{Proceedings of 1st international conference on image processing}, \bibinfo{organization}{IEEE}. pp. \bibinfo{pages}{168--172}.
%Type = Article
\bibitem[{Dabov et~al.(2007)Dabov, Foi, Katkovnik and Egiazarian}]{dabov2007image}
\bibinfo{author}{Dabov, K.}, \bibinfo{author}{Foi, A.}, \bibinfo{author}{Katkovnik, V.}, \bibinfo{author}{Egiazarian, K.}, \bibinfo{year}{2007}.
\newblock \bibinfo{title}{{Image denoising by sparse 3-D transform-domain collaborative filtering}}.
\newblock \bibinfo{journal}{IEEE Transactions on image processing} \bibinfo{volume}{16}, \bibinfo{pages}{2080--2095}.
%Type = Book
\bibitem[{Depeursinge et~al.(2017)Depeursinge, Omar, Al and Mitchell}]{depeursinge2017biomedical}
\bibinfo{author}{Depeursinge, A.}, \bibinfo{author}{Omar, S.}, \bibinfo{author}{Al, K.}, \bibinfo{author}{Mitchell, J.R.}, \bibinfo{year}{2017}.
\newblock \bibinfo{title}{{Biomedical texture analysis: fundamentals, tools and challenges}}.
\newblock \bibinfo{publisher}{Academic Press}.
%Type = Inproceedings
\bibitem[{Di~Feola et~al.(2023)Di~Feola, Tronchin and Soda}]{10178770}
\bibinfo{author}{Di~Feola, F.}, \bibinfo{author}{Tronchin, L.}, \bibinfo{author}{Soda, P.}, \bibinfo{year}{2023}.
\newblock \bibinfo{title}{A comparative study between paired and unpaired image quality assessment in low-dose ct denoising}, in: \bibinfo{booktitle}{{2023 IEEE 36th International Symposium on Computer-Based Medical Systems (CBMS)}}, pp. \bibinfo{pages}{471--476}.
\newblock \DOIprefix\doi{10.1109/CBMS58004.2023.00264}.
%Type = Article
\bibitem[{Du et~al.(2019)Du, Chen, Liao, Yang, Wang and Zhang}]{du2019visual}
\bibinfo{author}{Du, W.}, \bibinfo{author}{Chen, H.}, \bibinfo{author}{Liao, P.}, \bibinfo{author}{Yang, H.}, \bibinfo{author}{Wang, G.}, \bibinfo{author}{Zhang, Y.}, \bibinfo{year}{2019}.
\newblock \bibinfo{title}{{Visual attention network for low-dose CT}}.
\newblock \bibinfo{journal}{IEEE Signal Processing Letters} \bibinfo{volume}{26}, \bibinfo{pages}{1152--1156}.
%Type = Article
\bibitem[{Elad et~al.(2023)Elad, Kawar and Vaksman}]{elad2023image}
\bibinfo{author}{Elad, M.}, \bibinfo{author}{Kawar, B.}, \bibinfo{author}{Vaksman, G.}, \bibinfo{year}{2023}.
\newblock \bibinfo{title}{{Image denoising: The deep learning revolution and beyond—a survey paper}}.
\newblock \bibinfo{journal}{SIAM Journal on Imaging Sciences} \bibinfo{volume}{16}, \bibinfo{pages}{1594--1654}.
%Type = Article
\bibitem[{Gajera et~al.(2021)Gajera, Kapil, Ziaei, Mangalagiri, Siegel and Chapman}]{gajera2021ct}
\bibinfo{author}{Gajera, B.}, \bibinfo{author}{Kapil, S.R.}, \bibinfo{author}{Ziaei, D.}, \bibinfo{author}{Mangalagiri, J.}, \bibinfo{author}{Siegel, E.}, \bibinfo{author}{Chapman, D.}, \bibinfo{year}{2021}.
\newblock \bibinfo{title}{{CT-scan denoising using a charbonnier loss generative adversarial network}}.
\newblock \bibinfo{journal}{IEEE Access} \bibinfo{volume}{9}, \bibinfo{pages}{84093--84109}.
%Type = Article
\bibitem[{Gatys et~al.(2015)Gatys, Ecker and Bethge}]{gatys2015texture}
\bibinfo{author}{Gatys, L.}, \bibinfo{author}{Ecker, A.S.}, \bibinfo{author}{Bethge, M.}, \bibinfo{year}{2015}.
\newblock \bibinfo{title}{{Texture synthesis using convolutional neural networks}}.
\newblock \bibinfo{journal}{Advances in neural information processing systems} \bibinfo{volume}{28}.
%Type = Article
\bibitem[{Han et~al.(2021)Han, Shim and Baek}]{han2021low}
\bibinfo{author}{Han, M.}, \bibinfo{author}{Shim, H.}, \bibinfo{author}{Baek, J.}, \bibinfo{year}{2021}.
\newblock \bibinfo{title}{{Low-dose CT denoising via convolutional neural network with an observer loss function}}.
\newblock \bibinfo{journal}{Medical physics} \bibinfo{volume}{48}, \bibinfo{pages}{5727--5742}.
%Type = Article
\bibitem[{Han et~al.(2022)Han, Shim and Baek}]{han2022perceptual}
\bibinfo{author}{Han, M.}, \bibinfo{author}{Shim, H.}, \bibinfo{author}{Baek, J.}, \bibinfo{year}{2022}.
\newblock \bibinfo{title}{{Perceptual CT loss: implementing CT image specific perceptual loss for CNN-based low-dose CT denoiser}}.
\newblock \bibinfo{journal}{IEEE Access} \bibinfo{volume}{10}, \bibinfo{pages}{62412--62422}.
%Type = Article
\bibitem[{Huang et~al.(2021)Huang, Zhang, Zhang and Shan}]{huang2021gan}
\bibinfo{author}{Huang, Z.}, \bibinfo{author}{Zhang, J.}, \bibinfo{author}{Zhang, Y.}, \bibinfo{author}{Shan, H.}, \bibinfo{year}{2021}.
\newblock \bibinfo{title}{{DU-GAN: Generative adversarial networks with dual-domain U-Net-based discriminators for low-dose CT denoising}}.
\newblock \bibinfo{journal}{IEEE Transactions on Instrumentation and Measurement} \bibinfo{volume}{71}, \bibinfo{pages}{1--12}.
%Type = Inproceedings
\bibitem[{Isola et~al.(2017)Isola, Zhu, Zhou and Efros}]{isola2017image}
\bibinfo{author}{Isola, P.}, \bibinfo{author}{Zhu, J.Y.}, \bibinfo{author}{Zhou, T.}, \bibinfo{author}{Efros, A.A.}, \bibinfo{year}{2017}.
\newblock \bibinfo{title}{{Image-to-image translation with conditional adversarial networks}}, in: \bibinfo{booktitle}{Proceedings of the IEEE conference on computer vision and pattern recognition}, pp. \bibinfo{pages}{1125--1134}.
%Type = Article
\bibitem[{Izadi et~al.(2023)Izadi, Sutton and Hamarneh}]{izadi2023image}
\bibinfo{author}{Izadi, S.}, \bibinfo{author}{Sutton, D.}, \bibinfo{author}{Hamarneh, G.}, \bibinfo{year}{2023}.
\newblock \bibinfo{title}{{Image denoising in the deep learning era}}.
\newblock \bibinfo{journal}{Artificial Intelligence Review} \bibinfo{volume}{56}, \bibinfo{pages}{5929--5974}.
%Type = Inproceedings
\bibitem[{Johnson et~al.(2016)Johnson, Alahi and Fei-Fei}]{johnson2016perceptual}
\bibinfo{author}{Johnson, J.}, \bibinfo{author}{Alahi, A.}, \bibinfo{author}{Fei-Fei, L.}, \bibinfo{year}{2016}.
\newblock \bibinfo{title}{{Perceptual losses for real-time style transfer and super-resolution}}, in: \bibinfo{booktitle}{Computer Vision--ECCV 2016: 14th European Conference, Amsterdam, The Netherlands, October 11-14, 2016, Proceedings, Part II 14}, \bibinfo{organization}{Springer}. pp. \bibinfo{pages}{694--711}.
%Type = Article
\bibitem[{Kang et~al.(2023)Kang, Liu, Shu, Guo, Zhang, Zhou and Gui}]{kang2023gradient}
\bibinfo{author}{Kang, J.}, \bibinfo{author}{Liu, Y.}, \bibinfo{author}{Shu, H.}, \bibinfo{author}{Guo, N.}, \bibinfo{author}{Zhang, Q.}, \bibinfo{author}{Zhou, Y.}, \bibinfo{author}{Gui, Z.}, \bibinfo{year}{2023}.
\newblock \bibinfo{title}{{Gradient extraction based multiscale dense cross network for LDCT denoising}}.
\newblock \bibinfo{journal}{Nuclear Instruments and Methods in Physics Research Section A: Accelerators, Spectrometers, Detectors and Associated Equipment} \bibinfo{volume}{1055}, \bibinfo{pages}{168519}.
%Type = Article
\bibitem[{Kaur and Dong(2023)}]{kaur2023complete}
\bibinfo{author}{Kaur, A.}, \bibinfo{author}{Dong, G.}, \bibinfo{year}{2023}.
\newblock \bibinfo{title}{{A complete review on image denoising techniques for medical images}}.
\newblock \bibinfo{journal}{Neural Processing Letters} \bibinfo{volume}{55}, \bibinfo{pages}{7807--7850}.
%Type = Article
\bibitem[{Kingma and Ba(2014)}]{kingma2014adam}
\bibinfo{author}{Kingma, D.P.}, \bibinfo{author}{Ba, J.}, \bibinfo{year}{2014}.
\newblock \bibinfo{title}{{Adam: A method for stochastic optimization}}.
\newblock \bibinfo{journal}{arXiv preprint arXiv:1412.6980} .
%Type = Article
\bibitem[{Kwon and Ye(2021)}]{kwon2021cycle}
\bibinfo{author}{Kwon, T.}, \bibinfo{author}{Ye, J.C.}, \bibinfo{year}{2021}.
\newblock \bibinfo{title}{{Cycle-free cyclegan using invertible generator for unsupervised low-dose ct denoising}}.
\newblock \bibinfo{journal}{IEEE Transactions on Computational Imaging} \bibinfo{volume}{7}, \bibinfo{pages}{1354--1368}.
%Type = Inproceedings
\bibitem[{Kyung et~al.(2022)Kyung, Won, Pak, Hong and Kim}]{kyung2022mtd}
\bibinfo{author}{Kyung, S.}, \bibinfo{author}{Won, J.}, \bibinfo{author}{Pak, S.}, \bibinfo{author}{Hong, G.s.}, \bibinfo{author}{Kim, N.}, \bibinfo{year}{2022}.
\newblock \bibinfo{title}{{MTD-GAN: Multi-task Discriminator Based Generative Adversarial Networks for Low-Dose CT Denoising}}, in: \bibinfo{booktitle}{International Workshop on Machine Learning for Medical Image Reconstruction}, \bibinfo{organization}{Springer}. pp. \bibinfo{pages}{133--144}.
%Type = Article
\bibitem[{Li et~al.(2020a)Li, Hsu, Xie, Cong and Gao}]{li2020sacnn}
\bibinfo{author}{Li, M.}, \bibinfo{author}{Hsu, W.}, \bibinfo{author}{Xie, X.}, \bibinfo{author}{Cong, J.}, \bibinfo{author}{Gao, W.}, \bibinfo{year}{2020}a.
\newblock \bibinfo{title}{{SACNN: Self-attention convolutional neural network for low-dose CT denoising with self-supervised perceptual loss network}}.
\newblock \bibinfo{journal}{IEEE transactions on medical imaging} \bibinfo{volume}{39}, \bibinfo{pages}{2289--2301}.
%Type = Article
\bibitem[{Li et~al.(2022)Li, Li, Li, Wu, Zhao, Qiang and Tian}]{li2022adaptive}
\bibinfo{author}{Li, S.}, \bibinfo{author}{Li, Q.}, \bibinfo{author}{Li, R.}, \bibinfo{author}{Wu, W.}, \bibinfo{author}{Zhao, J.}, \bibinfo{author}{Qiang, Y.}, \bibinfo{author}{Tian, Y.}, \bibinfo{year}{2022}.
\newblock \bibinfo{title}{{An adaptive self-guided wavelet convolutional neural network with compound loss for low-dose CT denoising}}.
\newblock \bibinfo{journal}{Biomedical Signal Processing and Control} \bibinfo{volume}{75}, \bibinfo{pages}{103543}.
%Type = Article
\bibitem[{Li et~al.(2023)Li, Liu, Shu, Lu, Kang, Chen and Gui}]{li2023multi}
\bibinfo{author}{Li, Z.}, \bibinfo{author}{Liu, Y.}, \bibinfo{author}{Shu, H.}, \bibinfo{author}{Lu, J.}, \bibinfo{author}{Kang, J.}, \bibinfo{author}{Chen, Y.}, \bibinfo{author}{Gui, Z.}, \bibinfo{year}{2023}.
\newblock \bibinfo{title}{{Multi-Scale Feature Fusion Network for Low-Dose CT Denoising}}.
\newblock \bibinfo{journal}{Journal of Digital Imaging} , \bibinfo{pages}{1--18}.
%Type = Article
\bibitem[{Li et~al.(2021)Li, Shi, Xing, Miao, He, Yang and Jiang}]{li2021low}
\bibinfo{author}{Li, Z.}, \bibinfo{author}{Shi, W.}, \bibinfo{author}{Xing, Q.}, \bibinfo{author}{Miao, Y.}, \bibinfo{author}{He, W.}, \bibinfo{author}{Yang, H.}, \bibinfo{author}{Jiang, Z.}, \bibinfo{year}{2021}.
\newblock \bibinfo{title}{{Low-dose CT image denoising with improving WGAN and hybrid loss function}}.
\newblock \bibinfo{journal}{Computational and Mathematical Methods in Medicine} \bibinfo{volume}{2021}.
%Type = Article
\bibitem[{Li et~al.(2020b)Li, Zhou, Huang, Yu and Jin}]{li2020investigation}
\bibinfo{author}{Li, Z.}, \bibinfo{author}{Zhou, S.}, \bibinfo{author}{Huang, J.}, \bibinfo{author}{Yu, L.}, \bibinfo{author}{Jin, M.}, \bibinfo{year}{2020}b.
\newblock \bibinfo{title}{{Investigation of low-dose CT image denoising using unpaired deep learning methods}}.
\newblock \bibinfo{journal}{IEEE transactions on radiation and plasma medical sciences} \bibinfo{volume}{5}, \bibinfo{pages}{224--234}.
%Type = Article
\bibitem[{Liu et~al.(2017)Liu, Breuel and Kautz}]{liu2017unsupervised}
\bibinfo{author}{Liu, M.Y.}, \bibinfo{author}{Breuel, T.}, \bibinfo{author}{Kautz, J.}, \bibinfo{year}{2017}.
\newblock \bibinfo{title}{{Unsupervised image-to-image translation networks}}.
\newblock \bibinfo{journal}{Advances in neural information processing systems} \bibinfo{volume}{30}.
%Type = Article
\bibitem[{Ma et~al.(2020)Ma, Wei, Feng, He, Guo and Wang}]{ma2020low}
\bibinfo{author}{Ma, Y.}, \bibinfo{author}{Wei, B.}, \bibinfo{author}{Feng, P.}, \bibinfo{author}{He, P.}, \bibinfo{author}{Guo, X.}, \bibinfo{author}{Wang, G.}, \bibinfo{year}{2020}.
\newblock \bibinfo{title}{{Low-dose CT image denoising using a generative adversarial network with a hybrid loss function for noise learning}}.
\newblock \bibinfo{journal}{IEEE Access} \bibinfo{volume}{8}, \bibinfo{pages}{67519--67529}.
%Type = Inproceedings
\bibitem[{Marcos et~al.(2021)Marcos, Alirezaie and Babyn}]{marcos2021low}
\bibinfo{author}{Marcos, L.}, \bibinfo{author}{Alirezaie, J.}, \bibinfo{author}{Babyn, P.}, \bibinfo{year}{2021}.
\newblock \bibinfo{title}{{Low dose CT image denoising using boosting attention fusion gan with perceptual loss}}, in: \bibinfo{booktitle}{2021 43rd Annual International Conference of the IEEE Engineering in Medicine \& Biology Society (EMBC)}, \bibinfo{organization}{IEEE}. pp. \bibinfo{pages}{3407--3410}.
%Type = Inproceedings
\bibitem[{Matsoukas et~al.(2022)Matsoukas, Haslum, Sorkhei, S{\"o}derberg and Smith}]{matsoukas2022makes}
\bibinfo{author}{Matsoukas, C.}, \bibinfo{author}{Haslum, J.F.}, \bibinfo{author}{Sorkhei, M.}, \bibinfo{author}{S{\"o}derberg, M.}, \bibinfo{author}{Smith, K.}, \bibinfo{year}{2022}.
\newblock \bibinfo{title}{{What makes transfer learning work for medical images: Feature reuse \& other factors}}, in: \bibinfo{booktitle}{Proceedings of the IEEE/CVF Conference on Computer Vision and Pattern Recognition}, pp. \bibinfo{pages}{9225--9234}.
%Type = Article
\bibitem[{Mittal~Anish(2012)}]{mittal2012making}
\bibinfo{author}{Mittal~Anish, Rajiv~Soundararajan, A.C.B.}, \bibinfo{year}{2012}.
\newblock \bibinfo{title}{{Making a “completely blind” image quality analyzer}}.
\newblock \bibinfo{journal}{IEEE Signal processing letters} \bibinfo{volume}{20}, \bibinfo{pages}{209--212}.
%Type = Article
\bibitem[{Moen et~al.(2021)Moen, Chen, Holmes~III, Duan, Yu, Yu, Leng, Fletcher and McCollough}]{moen2021low}
\bibinfo{author}{Moen, T.R.}, \bibinfo{author}{Chen, B.}, \bibinfo{author}{Holmes~III, D.R.}, \bibinfo{author}{Duan, X.}, \bibinfo{author}{Yu, Z.}, \bibinfo{author}{Yu, L.}, \bibinfo{author}{Leng, S.}, \bibinfo{author}{Fletcher, J.G.}, \bibinfo{author}{McCollough, C.H.}, \bibinfo{year}{2021}.
\newblock \bibinfo{title}{{Low-dose CT image and projection dataset}}.
\newblock \bibinfo{journal}{Medical physics} \bibinfo{volume}{48}, \bibinfo{pages}{902--911}.
%Type = Article
\bibitem[{Ohno et~al.(2019)Ohno, Koyama, Seki, Kishida and Yoshikawa}]{ohno2019radiation}
\bibinfo{author}{Ohno, Y.}, \bibinfo{author}{Koyama, H.}, \bibinfo{author}{Seki, S.}, \bibinfo{author}{Kishida, Y.}, \bibinfo{author}{Yoshikawa, T.}, \bibinfo{year}{2019}.
\newblock \bibinfo{title}{{Radiation dose reduction techniques for chest CT: principles and clinical results}}.
\newblock \bibinfo{journal}{European journal of radiology} \bibinfo{volume}{111}, \bibinfo{pages}{93--103}.
%Type = Article
\bibitem[{Pan et~al.(2020)Pan, Yu, Wang, Xie, Sheng, Lei and Kwong}]{pan2020loss}
\bibinfo{author}{Pan, Z.}, \bibinfo{author}{Yu, W.}, \bibinfo{author}{Wang, B.}, \bibinfo{author}{Xie, H.}, \bibinfo{author}{Sheng, V.S.}, \bibinfo{author}{Lei, J.}, \bibinfo{author}{Kwong, S.}, \bibinfo{year}{2020}.
\newblock \bibinfo{title}{{Loss functions of generative adversarial networks (GANs): Opportunities and challenges}}.
\newblock \bibinfo{journal}{IEEE Transactions on Emerging Topics in Computational Intelligence} \bibinfo{volume}{4}, \bibinfo{pages}{500--522}.
%Type = Article
\bibitem[{Pang et~al.(2021)Pang, Lin, Qin and Chen}]{pang2021image}
\bibinfo{author}{Pang, Y.}, \bibinfo{author}{Lin, J.}, \bibinfo{author}{Qin, T.}, \bibinfo{author}{Chen, Z.}, \bibinfo{year}{2021}.
\newblock \bibinfo{title}{{Image-to-image translation: Methods and applications}}.
\newblock \bibinfo{journal}{IEEE Transactions on Multimedia} \bibinfo{volume}{24}, \bibinfo{pages}{3859--3881}.
%Type = Article
\bibitem[{Park et~al.(2019)Park, Baek, You, Choi and Seo}]{park2019unpaired}
\bibinfo{author}{Park, H.S.}, \bibinfo{author}{Baek, J.}, \bibinfo{author}{You, S.K.}, \bibinfo{author}{Choi, J.K.}, \bibinfo{author}{Seo, J.K.}, \bibinfo{year}{2019}.
\newblock \bibinfo{title}{{Unpaired image denoising using a generative adversarial network in X-ray CT}}.
\newblock \bibinfo{journal}{IEEE Access} \bibinfo{volume}{7}, \bibinfo{pages}{110414--110425}.
%Type = Article
\bibitem[{Park et~al.(2022)Park, Jeon, Lee and You}]{park2022denoising}
\bibinfo{author}{Park, H.S.}, \bibinfo{author}{Jeon, K.}, \bibinfo{author}{Lee, J.}, \bibinfo{author}{You, S.K.}, \bibinfo{year}{2022}.
\newblock \bibinfo{title}{{Denoising of pediatric low dose abdominal CT using deep learning based algorithm}}.
\newblock \bibinfo{journal}{Plos one} \bibinfo{volume}{17}, \bibinfo{pages}{e0260369}.
%Type = Article
\bibitem[{Russakovsky et~al.(2015)Russakovsky, Deng, Su, Krause, Satheesh, Ma, Huang, Karpathy, Khosla, Bernstein and Berg}]{russakovsky2015imagenet}
\bibinfo{author}{Russakovsky, O.}, \bibinfo{author}{Deng, J.}, \bibinfo{author}{Su, H.}, \bibinfo{author}{Krause, J.}, \bibinfo{author}{Satheesh, S.}, \bibinfo{author}{Ma, S.}, \bibinfo{author}{Huang, Z.}, \bibinfo{author}{Karpathy, A.}, \bibinfo{author}{Khosla, A.}, \bibinfo{author}{Bernstein, M.}, \bibinfo{author}{Berg, F.F.}, \bibinfo{year}{2015}.
\newblock \bibinfo{title}{{Imagenet large scale visual recognition challenge}}.
\newblock \bibinfo{journal}{International journal of computer vision} \bibinfo{volume}{115}, \bibinfo{pages}{211--252}.
%Type = Article
\bibitem[{Santucci et~al.(2021)Santucci, Faiella, Cordelli, Sicilia, de~Felice, Zobel, Iannello and Soda}]{santucci20213t}
\bibinfo{author}{Santucci, D.}, \bibinfo{author}{Faiella, E.}, \bibinfo{author}{Cordelli, E.}, \bibinfo{author}{Sicilia, R.}, \bibinfo{author}{de~Felice, C.}, \bibinfo{author}{Zobel, B.B.}, \bibinfo{author}{Iannello, G.}, \bibinfo{author}{Soda, P.}, \bibinfo{year}{2021}.
\newblock \bibinfo{title}{{3T MRI-radiomic approach to predict for lymph node status in breast cancer patients}}.
\newblock \bibinfo{journal}{Cancers} \bibinfo{volume}{13}, \bibinfo{pages}{2228}.
%Type = Book
\bibitem[{Scott(2015)}]{scott2015multivariate}
\bibinfo{author}{Scott, D.W.}, \bibinfo{year}{2015}.
\newblock \bibinfo{title}{{Multivariate density estimation: theory, practice, and visualization}}.
\newblock \bibinfo{publisher}{John Wiley \& Sons}.
%Type = Article
\bibitem[{Simonyan and Zisserman(2014)}]{simonyan2014very}
\bibinfo{author}{Simonyan, K.}, \bibinfo{author}{Zisserman, A.}, \bibinfo{year}{2014}.
\newblock \bibinfo{title}{{Very deep convolutional networks for large-scale image recognition}}.
\newblock \bibinfo{journal}{arXiv preprint arXiv:1409.1556} .
%Type = Article
\bibitem[{Tan et~al.(2022)Tan, Yang, You, Chen and Zhang}]{tan2022selective}
\bibinfo{author}{Tan, C.}, \bibinfo{author}{Yang, M.}, \bibinfo{author}{You, Z.}, \bibinfo{author}{Chen, H.}, \bibinfo{author}{Zhang, Y.}, \bibinfo{year}{2022}.
\newblock \bibinfo{title}{{A selective kernel-based cycle-consistent generative adversarial network for unpaired low-dose CT denoising}}.
\newblock \bibinfo{journal}{Precision Clinical Medicine} \bibinfo{volume}{5}, \bibinfo{pages}{pbac011}.
%Type = Inproceedings
\bibitem[{Tronchin et~al.(2021)Tronchin, Sicilia, Cordelli, Ramella and Soda}]{tronchin2021evaluating}
\bibinfo{author}{Tronchin, L.}, \bibinfo{author}{Sicilia, R.}, \bibinfo{author}{Cordelli, E.}, \bibinfo{author}{Ramella, S.}, \bibinfo{author}{Soda, P.}, \bibinfo{year}{2021}.
\newblock \bibinfo{title}{{Evaluating GANs in medical imaging}}, in: \bibinfo{booktitle}{Deep Generative Models, and Data Augmentation, Labelling, and Imperfections: First Workshop, MICCAI 2021, Proceedings}, \bibinfo{organization}{Springer}. pp. \bibinfo{pages}{112--121}.
%Type = Inproceedings
\bibitem[{Venkatanath et~al.(2015)Venkatanath, Praneeth, Bh, Channappayya and Medasani}]{venkatanath2015blind}
\bibinfo{author}{Venkatanath, N.}, \bibinfo{author}{Praneeth, D.}, \bibinfo{author}{Bh, M.C.}, \bibinfo{author}{Channappayya, S.S.}, \bibinfo{author}{Medasani, S.S.}, \bibinfo{year}{2015}.
\newblock \bibinfo{title}{{Blind image quality evaluation using perception based features}}, in: \bibinfo{booktitle}{2015 twenty first national conference on communications (NCC)}, \bibinfo{organization}{IEEE}. pp. \bibinfo{pages}{1--6}.
%Type = Misc
\bibitem[{{Vision and I. A. Group}(2009)}]{VisionIAGroup}
\bibinfo{author}{{Vision and I. A. Group}}, \bibinfo{year}{2009}.
\newblock \bibinfo{title}{{VIA/I-ELCAP public access research database}}.
\newblock \bibinfo{howpublished}{\url{http://www.via.cornell.edu/databases/lungdb.html}}.
\newblock \bibinfo{note}{Accessed April 7, 2009}.
%Type = Article
\bibitem[{Wang et~al.(2004)Wang, Bovik, Sheikh and Simoncelli}]{wang2004image}
\bibinfo{author}{Wang, Z.}, \bibinfo{author}{Bovik, A.C.}, \bibinfo{author}{Sheikh, H.R.}, \bibinfo{author}{Simoncelli, E.P.}, \bibinfo{year}{2004}.
\newblock \bibinfo{title}{Image quality assessment: from error visibility to structural similarity}.
\newblock \bibinfo{journal}{IEEE transactions on image processing} \bibinfo{volume}{13}, \bibinfo{pages}{600--612}.
%Type = Article
\bibitem[{Wolterink et~al.(2017)Wolterink, Leiner, Viergever and I{\v{s}}gum}]{wolterink2017generative}
\bibinfo{author}{Wolterink, J.M.}, \bibinfo{author}{Leiner, T.}, \bibinfo{author}{Viergever, M.A.}, \bibinfo{author}{I{\v{s}}gum, I.}, \bibinfo{year}{2017}.
\newblock \bibinfo{title}{{Generative adversarial networks for noise reduction in low-dose CT}}.
\newblock \bibinfo{journal}{IEEE transactions on medical imaging} \bibinfo{volume}{36}, \bibinfo{pages}{2536--2545}.
%Type = Inproceedings
\bibitem[{Yang et~al.(2023)Yang, Liu, Shang and Liu}]{yang2023adaptive}
\bibinfo{author}{Yang, L.}, \bibinfo{author}{Liu, H.}, \bibinfo{author}{Shang, F.}, \bibinfo{author}{Liu, Y.}, \bibinfo{year}{2023}.
\newblock \bibinfo{title}{{Adaptive Non-Local Generative Adversarial Networks for Low-Dose CT Image Denoising}}, in: \bibinfo{booktitle}{ICASSP 2023-2023 IEEE International Conference on Acoustics, Speech and Signal Processing (ICASSP)}, \bibinfo{organization}{IEEE}. pp. \bibinfo{pages}{1--5}.
%Type = Article
\bibitem[{Yang et~al.(2018)Yang, Yan, Zhang, Yu, Shi, Mou, Kalra, Zhang, Sun and Wang}]{yang2018low}
\bibinfo{author}{Yang, Q.}, \bibinfo{author}{Yan, P.}, \bibinfo{author}{Zhang, Y.}, \bibinfo{author}{Yu, H.}, \bibinfo{author}{Shi, Y.}, \bibinfo{author}{Mou, X.}, \bibinfo{author}{Kalra, M.K.}, \bibinfo{author}{Zhang, Y.}, \bibinfo{author}{Sun, L.}, \bibinfo{author}{Wang, G.}, \bibinfo{year}{2018}.
\newblock \bibinfo{title}{{Low-dose CT image denoising using a generative adversarial network with Wasserstein distance and perceptual loss}}.
\newblock \bibinfo{journal}{IEEE transactions on medical imaging} \bibinfo{volume}{37}, \bibinfo{pages}{1348--1357}.
%Type = Article
\bibitem[{Yin et~al.(2021)Yin, Xia, He, Zhang, Wang and Zu}]{yin2021unpaired}
\bibinfo{author}{Yin, Z.}, \bibinfo{author}{Xia, K.}, \bibinfo{author}{He, Z.}, \bibinfo{author}{Zhang, J.}, \bibinfo{author}{Wang, S.}, \bibinfo{author}{Zu, B.}, \bibinfo{year}{2021}.
\newblock \bibinfo{title}{{Unpaired image denoising via Wasserstein GAN in low-dose CT image with multi-perceptual loss and fidelity loss}}.
\newblock \bibinfo{journal}{Symmetry} \bibinfo{volume}{13}, \bibinfo{pages}{126}.
%Type = Article
\bibitem[{Yin et~al.(2023)Yin, Xia, Wang, He, Zhang and Zu}]{yin2023unpaired}
\bibinfo{author}{Yin, Z.}, \bibinfo{author}{Xia, K.}, \bibinfo{author}{Wang, S.}, \bibinfo{author}{He, Z.}, \bibinfo{author}{Zhang, J.}, \bibinfo{author}{Zu, B.}, \bibinfo{year}{2023}.
\newblock \bibinfo{title}{{Unpaired low-dose CT denoising via an improved cycle-consistent adversarial network with attention ensemble}}.
\newblock \bibinfo{journal}{The Visual Computer} \bibinfo{volume}{39}, \bibinfo{pages}{4423--4444}.
%Type = Article
\bibitem[{You et~al.(2018)You, Yang, Shan, Gjesteby, Li, Ju, Zhang, Zhao, Zhang, Cong and Weng}]{you2018structurally}
\bibinfo{author}{You, C.}, \bibinfo{author}{Yang, Q.}, \bibinfo{author}{Shan, H.}, \bibinfo{author}{Gjesteby, L.}, \bibinfo{author}{Li, G.}, \bibinfo{author}{Ju, S.}, \bibinfo{author}{Zhang, Z.}, \bibinfo{author}{Zhao, Z.}, \bibinfo{author}{Zhang, Y.}, \bibinfo{author}{Cong, W.}, \bibinfo{author}{Weng}, \bibinfo{year}{2018}.
\newblock \bibinfo{title}{{Structurally-sensitive multi-scale deep neural network for low-dose CT denoising}}.
\newblock \bibinfo{journal}{IEEE access} \bibinfo{volume}{6}, \bibinfo{pages}{41839--41855}.
%Type = Inproceedings
\bibitem[{Zamir et~al.(2021)Zamir, Arora, Khan, Hayat, Khan, Yang and Shao}]{zamir2021multi}
\bibinfo{author}{Zamir, S.W.}, \bibinfo{author}{Arora, A.}, \bibinfo{author}{Khan, S.}, \bibinfo{author}{Hayat, M.}, \bibinfo{author}{Khan, F.S.}, \bibinfo{author}{Yang, M.H.}, \bibinfo{author}{Shao, L.}, \bibinfo{year}{2021}.
\newblock \bibinfo{title}{{Multi-stage progressive image restoration}}, in: \bibinfo{booktitle}{Proceedings of the IEEE/CVF conference on computer vision and pattern recognition}, pp. \bibinfo{pages}{14821--14831}.
%Type = Inproceedings
\bibitem[{Zhang et~al.(2019)Zhang, Goodfellow, Metaxas and Odena}]{zhang2019self}
\bibinfo{author}{Zhang, H.}, \bibinfo{author}{Goodfellow, I.}, \bibinfo{author}{Metaxas, D.}, \bibinfo{author}{Odena, A.}, \bibinfo{year}{2019}.
\newblock \bibinfo{title}{{Self-attention generative adversarial networks}}, in: \bibinfo{booktitle}{International conference on machine learning}, \bibinfo{organization}{PMLR}. pp. \bibinfo{pages}{7354--7363}.
%Type = Inproceedings
\bibitem[{Zhu et~al.(2017)Zhu, Park, Isola and Efros}]{zhu2017unpaired}
\bibinfo{author}{Zhu, J.Y.}, \bibinfo{author}{Park, T.}, \bibinfo{author}{Isola, P.}, \bibinfo{author}{Efros, A.A.}, \bibinfo{year}{2017}.
\newblock \bibinfo{title}{{Unpaired image-to-image translation using cycle-consistent adversarial networks}}, in: \bibinfo{booktitle}{Proceedings of the IEEE international conference on computer vision}, pp. \bibinfo{pages}{2223--2232}.

\end{thebibliography}

\end{document}

%\begin{frontmatter}
%\begin{titlepage}
\begin{center}
\vspace*{0.2cm}
\large\textbf{Supplementary material}
\vspace{0.5cm}
\noindent\rule{14cm}{0.4pt}
\Large\textbf{Multi-Scale Texture Loss for CT Denoising with GANs}

\vspace{1cm}

\small
Francesco Di Feola \textsuperscript{a}, Lorenzo Tronchin \textsuperscript{b}, Valerio Guarrasi,\textsuperscript{b}
Paolo Soda\textsuperscript{a, b}
\scriptsize

\textit{\textsuperscript{a}Department of Diagnostics and Intervention, Radiation Physics, Biomedical Engineering, Umeå University, Sweden}

\textit{\textsuperscript{b}Unit of Computer Systems and Bioinformatics, Department of Engineering, University Campus Bio-Medico of Rome, Italy}

\end{center}

%\end{titlepage}

\normalsize

%%%%%%%%%%%%%%%%%%%%%%%%%%%%%%%%%%%%%%%%%%%%%%%%%%%%%%%%%%%%%%%%%%%%%%%%%%%%%
%%%%%%%%%%%%%%%%%%%%%%%%%%%%%%%%%%%%%%%%%%%%%%%%%%%%%%%%%%%%%%%%%%%%%%%%%%%%%
\section{Introduction}

This document provides supplementary material to the main manuscript, providing a detailed description of the GAN backbones used, figures, and tables, there omitted for readability.
In detail, this document is organized as follows: 
section~\ref{suppl:architectures} reports a description of the GAN backbones used, providing information about their architectures as well as the computation of their loss function; section~\ref{suppl:ids} provides the list of patient IDs across all the datasets described in the main manuscript; section~\ref{suppl:stats} presents statistical analysis performed on a per-patient basis; section~\ref{suppl:visuals} presents visual results attained using CycleGAN, Pix2Pix and UNIT across various configurations of the loss function;

%%%%%%%%%%%%%%%%%%%%%%%%%%%%%%%%%%%%%%%%%%%%%%%%%%%%%%%%%%%%%%%%%%%%%%%%%%%%%
%%%%%%%%%%%%%%%%%%%%%%%%%%%%%%%%%%%%%%%%%%%%%%%%%%%%%%%%%%%%%%%%%%%%%%%%%%%%%
\section{Deep Models} \label{suppl:architectures}
This section describes the GAN architectures used in this work. Let us remember that $\bm{x} \in X$ is a LDCT image, whilst $\bm{y} \in Y$ is an HDCT image.
\subsubsection{Pix2Pix}
Pix2Pix is a well-established baseline for image-to-image translation tasks~\citep{pang2021image}.
Based on conditional GAN (cGAN), Pix2Pix requires pairs of images in the training phase, restricting its application to paired datasets. 
It uses a generator $G$ and a discriminator $D$ which, in the original paper~\citep{isola2017image}, are implemented as a U-Net encoder-decoder architecture and a PatchGAN, respectively.
The loss function of a cGAN is the adversarial loss:
\begin{equation}
  \mathcal{L}_{cGAN}(G,D) = \mathbb{E}_{\bm{x},\bm{y}}[\log D(\bm{x}, \bm{y})] + \mathbb{E}_{\bm{x}, \bm{z}}[ \log(1-D(G(\bm{x}, \bm{z}))]  
\end{equation}
where $G$ and $D$ are trained in an alternating fashion performing a min-max optimization, $\mathbb{E}$ is the expectation operator, and $\bm{z}$ is a Gaussian noise vector that in~\citep{isola2017image} is provided in the form of dropout applied on the layers of $G$.
In Pix2Pix, the goal of the generator is not only to fool the discriminator but also to generate $G(\bm{x}, \bm{z})\approx \bm{y}$. 

To this end, a pixel-wise regression loss is used:
\begin{equation}
\mathcal{L}_{L1}(G) = \mathbb{E}_{\bm{x},\bm{y},\bm{z}}[\ ||\bm{y}- G(\bm{x},\bm{z})||_{1}]\
\end{equation}
where $||\cdot||_{1}$ is the L-1 distance.
The final objective function of Pix2Pix is:
%\begin{equation}
%\mathcal{L}(G,D)=\mathcal{L}_{cGAN}%(G,D)+\lambda\mathcal{L}_{1}(G) 
%\end{equation}
\begin{equation}
\mathcal{L}=\mathcal{L}_{cGAN}+\lambda\mathcal{L}_{1} 
\label{eq:pix2pix}
\end{equation}
where $\lambda$ is a weighting parameter that is empirically set to 1~\citep{isola2017image}.
  
\subsubsection{CycleGAN}
Paired translation is not very practical and feasible, especially in the medical domain where the acquisition of paired datasets may be complex. 
For this reason, unpaired methods have attracted more attention and CycleGAN established itself as a milestone architecture. 
Being a bidirectional generative model that contains two generators ($G$, $F$) and two discriminators ($D_{X}$, $D_{Y}$), CycleGAN tackles unpaired translation introducing the cycle-consistency constraint which attempts to regularize and reduce the space of possible mappings~\citep{zhu2017unpaired}, helping $G$ and $F$ to be bijective functions. 
In the original paper~\citep{zhu2017unpaired}, $G$ and $F$ are implemented as residual encoder-decoder architectures, whilst $D_{X}$ and $D_{Y}$ are implemented as PatchGANs.
CycleGAN applies least squares adversarial losses to both generators. 
For the generator $G$ and the discriminator $D_{Y}$ the adversarial loss is expressed as:
\begin{equation}
\mathcal{L}_{LSGAN}(G,D_{Y},X,Y) = \mathbb{E}_{\bm{y}}[\ (D_{Y}(\bm{y})-1)^2]\ +  \mathbb{E}_{\bm{x}}[\ (D_{Y}(G(\bm{x}))^2].\
\end{equation}
A similar adversarial loss is introduced for the generator $F$ and the discriminator $D_{X}$.

The cycle consistency constraint is implemented as a loss term called cycle consistency loss:
\begin{equation}
\mathcal{L}_{cyc}(G, F) = \mathbb{E}_{\bm{x}}[\ ||F(G(\bm{x}))- \bm{x}||_{1}]\ +\mathbb{E}_{\bm{y}}[\ ||G(F(\bm{y}))- \bm{y}||_{1}]\
\end{equation}
which encourages $F(G(\bm{x}))=\Tilde{\bm{x}} \approx \bm{x}$ and $G(F(\bm{y}))=\Tilde{\bm{y}} \approx \bm{y}$, i.e., it imposes latent space constraints to encourage the information content to be preserved during the reconstruction mapping.

Similarly to the cycle consistency loss, an identity loss function is introduced to further regularize the mapping behavior of the generators:
\begin{equation}
\mathcal{L}_{idt}(G, F) = \mathbb{E}_{\bm{x}}[\ ||F(\bm{x})- \bm{x}||_{1}]\
+\mathbb{E}_{\bm{y}}[\ ||G(\bm{y})- \bm{y}||_{1}].\
\end{equation}
The final objective function of CycleGAN is:
%\begin{equation}
%\begin{split}
%\mathcal{L}(G,F,D_{X}, D_{Y}) = \mathcal{L}_{GAN}%(G,D_{Y}, X, Y)+\\
%+\mathcal{L}_{GAN}(F, D_{X}, X, Y) + %\lambda_{1}\mathcal{L}_{cyc}(G, F)+\\
% + \lambda_{2}\mathcal{L}_{idt}(G,F)
%\end{split}
%\end{equation}
\begin{equation}
\begin{split}
\mathcal{L} = \mathcal{L}_{SGAN}
+\lambda_{1}\mathcal{L}_{cyc}
 + \lambda_{2}\mathcal{L}_{idt}
\end{split}
\label{eq:cycleGAN}
\end{equation}
where the weighting parameters $\lambda_{1}$ and $\lambda_{2}$ are empirically set to 10 and 0.5~\citep{zhu2017unpaired}, respectively.

\subsubsection{UNIT}
UNIT (UNsupervised Image-to-Image-Translation) tackles unpaired scenarios by assuming that a pair of corresponding images in different domains can be mapped to the same representation in a shared latent space~\citep{pang2021image}. 
UNIT is composed of six sub-networks: two encoders $E_{X}$, $E_{Y}$ that map the input to a shared latent representation, two decoders $G_{X}$, $G_{Y}$ that reconstruct the translated output and two discriminators $D_{X}$ and $D_{Y}$. 
In the original paper~\citep{liu2017unsupervised}, $E_{X}$, $E_{Y}$, $G_{X}$, $G_{Y}$ share the same structure consisting of 3 convolutional layers and 4 basic residual blocks, whilst $D_{X}$, $D_{Y}$ are implemented as PatchGANs.
The optimization problem is solved by interpreting the roles of the six subnetworks as follows: the encoder-decoder pairs $\{E_{X}, G_{X}\}$ and  $\{E_{Y}, G_{Y}\}$ constitute two variational autoencoders referred as $VAE_{X}$ and $VAE_{Y}$, whilst $\{G_{X}, D_{X}\}$ and $\{G_{Y}, D_{Y}\}$ are two GANs referred as $GAN_{X}$ and $GAN_{Y}$.

The adversarial losses are:
\begin{equation}
\mathcal{L}_{GAN_{X}}(E_{Y}, G_{X}, D_{X}) = \lambda_{0}\mathbb{E}_{\bm{x}}[\ \log D_{X}(\bm{x})]\ + \lambda_{0}\mathbb{E}_{\bm{z_{Y}}}[\ \log(1-D_{1}(G_{1}(\bm{z_{Y}}))]\
\end{equation}

\begin{equation}
\mathcal{L}_{GAN_{Y}}(E_{X}, G_{Y}, D_{Y}) = \lambda_{0}\mathbb{E}_{\bm{y}}[\ \log D_{Y}(\bm{y})]\ +  \lambda_{0}\mathbb{E}_{\bm{z_{X}}}[\ \log(1-D_{Y}(G_{Y}(\bm{z_{X}}))]\
\end{equation}

where $\bm{z_{X}}$ and $\bm{z_{Y}}$ are the latent representations produced by the two encoders $E_{X}$ and $E_{Y}$, respectively.

$VAE_{X}$ and $VAE_{Y}$ are trained to minimize a variational upper bound with their loss function defined as: 
\begin{equation}
\mathcal{L}_{VAE_{X}}(E_{X}, G_{X}) = \lambda_{1}KL(q_{X}(\bm{z_{X}}|\bm{x})||p_{\eta}(\bm{z})) 
- \lambda_{2}\mathbb{E}_{\bm{z_{X}}} [\log( p_{G_{X}}(\bm{x}|\bm{z_{X}}))]\
\end{equation}

\begin{equation}
\mathcal{L}_{VAE_{Y}}(E_{Y}, G_{Y}) = \lambda_{1}KL(q_{Y}(\bm{z_{Y}}|\bm{y})||p_{\eta}(\bm{z}))
- \lambda_{2}\mathbb{E}_{\bm{z_{Y}}} [\log( p_{G_{Y}}(\bm{y}|\bm{z_{Y}}))]\
\end{equation}
where $KL$ is the Kullback Leibler divergence that penalizes deviation of the distribution of the latent codes  $q_{1}(\bm{z_{X}}|\bm{x})$, $q_{2}(\bm{z_{Y}}|\bm{y})$ from the prior distribution $p_{\eta}(\bm{z})$, whilst $p_{G_{X}}(\bm{x}|\bm{z_{X}})$ and $p_{G_{Y}}(\bm{y}|\bm{z_{Y}})$ are distributions of the reconstructed samples starting from the latent codes $\bm{z_{X}}$, $\bm{z_{Y}}$. 
On a practical level, the negative likelihood can be replaced with the $L$-1 distance computed between an image and its reconstruction, i.e., $||\bm{x}-G_{X}(\bm{z_{X}})||_{1}$ or $||\bm{y}-G_{Y}(\bm{z_{Y}})||_{1}$.
The weights $\lambda_{1}$, $\lambda_{2}$ are set to default values, 0.1 and 100, respectively.

A similar term is used to model the cycle-consistency constraint, given by:

\begin{equation}
\begin{split}
\mathcal{L}_{cyc_{X}}(E_{X}, G_{X}, E_{Y}, G_{Y}) = \lambda_{3}KL(q_{X}(\bm{\hat{z}_{X}}|\bm{x})||p_{\eta}( \bm{z})) 
+\lambda_{3}KL(q_{Y}(\bm{\hat{z}_{Y}}|\bm{\hat{y}})||p_{\eta}(\bm{z}))+\\
- \lambda_{4}\mathbb{E}_{\bm{\hat{z}_{Y}}} [\log( p_{G_{X}}(\bm{x}|\bm{\hat{z}_{Y}}))]\
\end{split}
\end{equation}

\begin{equation}
\begin{split}
\mathcal{L}_{cyc_{Y}}(E_{Y}, G_{Y}, E_{X}, G_{X}) = \lambda_{3}KL(q_{Y}(\bm{\hat{z}_{Y}}|\bm{y})||p_{\eta}(\bm{z})) 
+\lambda_{3}KL(q_{X}(\bm{\hat{z}_{X}}|\bm{\hat{x}})||p_{\eta}(\bm{z}))+\\
- \lambda_{4}\mathbb{E}_{\bm{\hat{z}_{X}}} [\log( p_{G_{Y}}(\bm{y}|\bm{\hat{z}_{X}}))]\
\end{split}
\end{equation}
where the KL terms penalize the latent codes $q_{X}(\bm{z_{X}}|\bm{x})$, $q_{Y}(\bm{z_{Y}}|\bm{\hat{y}})$, $q_{Y}(\bm{z_{Y}}|\bm{y})$ and $q_{X}(\bm{z_{X}}|\bm{\hat{x}})$ deviating from the prior distribution $p_{\eta}(\bm{z})$. 
The generated samples are $\bm{\hat{x}}=G_{X}(\bm{z_{Y}})$ and $\bm{\hat{y}}=G_{Y}(\bm{z_{X}})$ whilst $\bm{\hat{z}_{X}}=E_{X}(\bm{\hat{x}})$ and $\bm{\hat{z}_{Y}}=E_{Y}(\bm{\hat{y}})$ are latent representations. Also here, the negative log-likelihoods can be expressed in terms of L1-distances, i.e., $||\bm{x}-G_{X}(\bm{\hat{z}_{X}})||_{1}$ and  $||\bm{y}-G_{Y}(\bm{\hat{z}_{Y}})||_{1}$ where $G_{X}(\bm{\hat{z}_{X}})=\bm{\Tilde{x}}$ and $G_{Y}(\bm{\hat{z}_{Y}})=\bm{\Tilde{y}}$.
The weights of each loss term are set to default values with $\lambda_{0}=10$,  $\lambda_{1}=\lambda_{3}=0.1$ and $\lambda_{2}=\lambda_{4}=100$.

The final objective function of UNIT is computed as:
\begin{equation}
\mathcal{L} = \mathcal{L}_{VAE}
+\mathcal{L}_{GAN}+ \mathcal{L}_{cyc}.
\label{eq:UNIT}
\end{equation}

\section{Patients' IDs} \label{suppl:ids}
LDCT-and-Projection Data by Mayo Clinic (\cite{moen2021low}):
\begin{itemize}
    \item Training set: L096, L109, L143, L192, L286, L291, L310, L333, L506;
    \item Test set (referred to as \emph{Dataset A} in this manuscript): L067, C002, C004, C030, C067, C120, C124, C166;
\end{itemize}

LIDC/IDRI (referred to as \emph{Dataset B} in this manuscript) (\cite{armato2011lung}): 

0361, 0369, 0555, 0574, 0785, 0928, 0930, 0959.

ELCAP dataset (referred to as \emph{Dataset C} in this manuscript) (\cite{VisionIAGroup}): W0001, W0002, W0003, W0004, W0005, W0006, W0007, W0008, W0009, W0010, 

W0011, W0012, W0013, W0014, W0015, W0016, W0017, W0018, W0019, W0020, 

W0021, W0022, W0023, W0024, W0025, W0026, W0027, W0028, W0029, W0030, 

W0031, W0032, W0033, W0034, W0035, W0036, W0037, W0038, W0039, W0040, 

W0041, W0042, W0043, W0044, W0045, W0046, W0047,W0048, W0049, W0050.

\section{Statistical analysis} \label{suppl:stats}
This section presents the statistical analysis conducted on an individual patient basis. We evaluate the statistical significance of the best-performing loss function configuration compared to other configurations using the Wilcoxon signed rank test, with a p-value threshold  set at $p\leq0.05$. 
The main manuscript demonstrates that MSTLF attains the best overall performance across different datasets and GAN architectures. 
In particular, MSTLF-average yields the most favorable results on paired metrics (PSNR, MSE, and SSIM), whilst MSTLF-attention excels in unpaired metrics (NIQE and PIQUE). 

Hereinafter, the  tables share the same structure reporting the following information: the dataset used (first row), GAN architecture and performance metrics (second row), 
the configuration of the loss function whose statistical significance is assessed against the other configurations (first and second columns, respectively). 
For each comparison, the tables report the number of patients for whom the loss function in the first column is statistically better than the loss function in the second column. 
The second-to-last row and column sum up   the values to summarize the results across metrics and loss functions configurations, while the last row and column show the percentage.

The per-patient analysis offers additional insights into the findings presented in the main manuscript, confirming the results attained. 
In particular,~\tablename~\ref{tab:dataset_A_pix2pix}-\ref{tab:dataset_C_UNIT} corroborate the results reported in Tables 3 and 4 of the main document. 
It is worth noting that although we showed the agnosticism of the proposed MSTLF with respect to the GAN architecture used, the per-patient analysis brings out that among the three architectures, cycleGAN is the one with the largest percentage in the last column (\tablename~\ref{tab:dataset_A_cycle}, \ref{tab:dataset_A_cycle_un}, \ref{tab:dataset_B_cycle}, \ref{tab:dataset_C_cycle}). 
%Such result does not alter our findings since it testifies differences in training stability between different architectures that are beyond the scope of this paper and lends itself to future investigation.
%~\ref{tab:dataset_A_pix2pix_un},~\ref{tab:dataset_A_cycle_un},~\ref{tab:dataset_A_UNIT_un} 

%Green denotes that the best-performing configuration is %also statistically better with respect to another %configuration on an individual-patient basis; red denotes %that the best-performing configuration is statistically %worse with respect to another configuration on an %individual-patient basis; yellow denotes that the best-%performing configuration is not statistically different %from another configuration on an individual patient basis. 

\begin{table}[h]
\centering
\resizebox{.75\textwidth}{!}{
\begin{tabular}{|c|c|c|c|c|c|c|}
\hline
\multicolumn{7}{|c|}{\emph{Dataset A}}\\
\hline
\multicolumn{2}{|c|}{\textbf{Pix2Pix}} & \textbf{PNSR}& \textbf{MSE}& \textbf{SSIM} & \textbf{Total} & \textbf{\%} \\
\cline{1-7}
\multirow{10}{9em}{MSTLF-average vs}
&Baseline& 8 & 6 & 8 & 22/24 & 92$\%$\\
\cline{2-7}
&VGG-16 & 8 & 7 & 8 & 23/24 & 96$\%$\\
\cline{2-7}
&AE-CT  & 6 & 7 & 6 & 19/24 & 79$\%$\\
\cline{2-7}
&SSIM-L & 3 & 6 & 0 & 9/24 & 38$\%$\\
\cline{2-7}
&EDGE & 6 & 7 & 8 & 21/24 & 88$\%$\\
\cline{2-7}
&MSTLF-max & 8 & 8 & 8 & 24/24 & 100$\%$\\
\cline{2-7}
&MSTLF-Frobenius & 8 & 8 & 7 & 23/24 & 96$\%$\\
\cline{2-7}
&MSTLF-attention & 8 & 8 & 8 & 24/24 & 100$\%$\\
\cline{2-7}
&\textbf{Total} & 55/64 & 57/64 & 53/64  \\
\cline{2-5}
&\textbf{\%} & 86$\%$ & 89$\%$ & 83$\%$ \\
\cline{1-5}
\end{tabular}}
\caption{Wilcoxon signed rank test conducted on a per patient basis within \emph{Dataset A} (Mayo simulated data, 8 patients in total) to assess the statistical significance of MSTLF-average in comparison to other loss function configurations, with Pix2Pix. 
Each box shows the number of patients for whom MSTLF-average is statistically better than other loss functions.}
\label{tab:dataset_A_pix2pix}
\end{table}

\begin{table}[h]
\centering
\resizebox{.75\textwidth}{!}{
\begin{tabular}{|c|c|c|c|c|c|c|}
\hline
\multicolumn{7}{|c|}{\emph{Dataset A}}\\
\hline
%\multicolumn{2}{c|}{OBJECTIVE FUNCTION}
\multicolumn{2}{|c|}{\textbf{CycleGAN}} & \textbf{PNSR}& \textbf{MSE}& \textbf{SSIM} & \textbf{Total} & \textbf{\%} \\
\cline{1-7}
\multirow{10}{9em}{MSTLF-average vs}
&Baseline& 8 & 8 & 8 & 24/24 & 100$\%$\\
\cline{2-7}
&VGG-16 & 7 & 7 & 6 & 20/24 & 83$\%$\\
\cline{2-7}
&AE-CT  & 4 & 4 & 7 & 15/24 & 63$\%$\\
\cline{2-7}
&SSIM-L & 8 & 8 & 8 & 24/24 & 100$\%$\\
\cline{2-7}
&EDGE & 8 & 8 & 8 & 24/24 & 100$\%$\\
\cline{2-7}
&MSTLF-max & 7 & 7 & 7 & 21/24 & 88$\%$\\
\cline{2-7}
&MSTLF-Frobenius & 6 & 6 & 6 & 18/24 & 75$\%$\\
\cline{2-7}
&MSTLF-attention & 7 & 7 & 6 & 20/24 & 83$\%$\\
\cline{2-7}
&\textbf{Total} & 55/64 & 57/64 & 56/64\\
\cline{2-5}
&\textbf{\%} & 86$\%$ & 86$\%$ & 88$\%$\\
\cline{1-5}
\end{tabular}}
\caption{Wilcoxon signed rank test conducted on a per patient basis within \emph{Dataset A} (Mayo simulated data, 8 patients in total) to assess the statistical significance of MSTLF-average in comparison to other loss function configurations in CycleGAN. Each box shows the number of patients for whom MSTLF-average is statistically better.}
\label{tab:dataset_A_cycle}
\end{table}

\begin{table}[h]
\centering
\resizebox{.75\textwidth}{!}{
\begin{tabular}{|c|c|c|c|c|c|c|}
\hline
\multicolumn{7}{|c|}{\emph{Dataset A}}\\
\hline
%\multicolumn{2}{c|}{OBJECTIVE FUNCTION}
\multicolumn{2}{|c|}{\textbf{UNIT}}& \textbf{PNSR}& \textbf{MSE}& \textbf{SSIM} & \textbf{Total} & \textbf{\%} \\
\cline{1-7}
\multirow{10}{9em}{MSTLF-average vs}
&Baseline & 8 & 8 & 8 & 24/24 & 100$\%$\\
\cline{2-7}
&VGG-16 & 8 & 8 & 8 & 24/24 & 100$\%$\\
\cline{2-7}
&AE-CT  & 8 & 8 & 8 & 24/24 & 100$\%$\\
\cline{2-7}
&SSIM-L & 7 & 7 & 8 & 22/24 & 92$\%$\\
\cline{2-7}
&EDGE & 8 & 8 & 8 & 24/24 & 100$\%$\\
\cline{2-7}
&MSTLF-max & 8 & 8 & 8 & 24/24 & 100$\%$\\
\cline{2-7}
&MSTLF-Frobenius & 8 & 8 & 8 & 24/24 & 100$\%$\\
\cline{2-7}
&MSTLF-attention & 6 & 5 & 8 & 19/24 & 79$\%$\\
\cline{2-7}
&\textbf{Total} & 61/64 & 60/64 & 64/64\\
\cline{2-5}
&\textbf{\%} & 92$\%$ & 94$\%$ & 100$\%$\\
\cline{1-5}
\end{tabular}}
\caption{Wilcoxon signed rank test conducted on a per patient basis within \emph{Dataset A} (Mayo simulated data, 8 patients in total) to assess the statistical significance of MSTLF-average in comparison to other loss function configurations in UNIT. Each box shows the number of patients for whom MSTLF-average is statistically better.}
\label{tab:dataset_A_UNIT}
\end{table}

\begin{table}[h]
\centering
\resizebox{.7\textwidth}{!}{
\begin{tabular}{|c|c|c|c|c|c|}
\hline
\multicolumn{6}{|c|}{\emph{Dataset A}}\\
\hline
%\multicolumn{2}{c|}{OBJECTIVE FUNCTION}
\multicolumn{2}{|c|}{\textbf{Pix2Pix}} & \textbf{NIQE}& \textbf{PIQUE} & \textbf{Total} & \textbf{\%} \\
\cline{1-6}
\multirow{10}{9em}{MSTLF-attention vs}
&Baseline & 8 & 7 & 15/16 & 94$\%$\\
\cline{2-6}
&VGG-16 & 8 & 7 & 15/16 & 94$\%$\\
\cline{2-6}
&AE-CT  & 8 & 7 & 15/16 & 94$\%$\\
\cline{2-6}
&SSIM-L & 8 & 7 & 15/16 & 94$\%$\\
\cline{2-6}
&EDGE & 8 & 7 & 15/16 & 94$\%$\\
\cline{2-6}
&MSTLF-max & 8 & 7 & 15/16 & 94$\%$\\
\cline{2-6}
&MSTLF-average & 8 & 7 & 15/16 & 94$\%$\\
\cline{2-6}
&MSTLF-Frobenius & 8 & 7 & 15/16 & 94$\%$\\
\cline{2-6}
&\textbf{Total} & 64/64 & 56/64\\
\cline{2-4}
&\textbf{\%} & 100$\%$ & 88$\%$\\
\cline{1-4}
\end{tabular}}
\caption{Wilcoxon signed rank test conducted on a per patient basis within \emph{Dataset A} (Mayo simulated data, 8 patients in total) to assess the statistical significance of MSTLF-attention in comparison to other loss function configurations in Pix2Pix. Each box shows the number of patients for whom MSTLF-attention is statistically better.}
\label{tab:dataset_A_pix2pix_un}
\end{table}

\begin{table}[h]
\centering
\resizebox{.7\textwidth}{!}{
\begin{tabular}{|c|c|c|c|c|c|}
\hline
\multicolumn{6}{|c|}{\emph{Dataset A}}\\
\hline
%\multicolumn{2}{c|}{OBJECTIVE FUNCTION}
\multicolumn{2}{|c|}{\textbf{CycleGAN}} & \textbf{NIQE}& \textbf{PIQUE} & \textbf{Total} & \textbf{\%} \\
\cline{1-6}
\multirow{10}{9em}{MSTLF-attention vs}
&Baseline & 8 & 8 & 16/16 & 100$\%$\\
\cline{2-6}
&VGG-16 & 8 & 8 & 16/16 & 100$\%$\\
\cline{2-6}
&AE-CT  & 8 & 8 & 16/16 & 100$\%$\\
\cline{2-6}
&SSIM-L & 8 & 8 & 16/16 & 100$\%$\\
\cline{2-6}
&EDGE & 8 & 8 & 16/16 & 100$\%$\\
\cline{2-6}
&MSTLF-max & 8 & 8 & 16/16 & 100$\%$\\
\cline{2-6}
&MSTLF-average & 8 & 8 & 16/16 & 100$\%$\\
\cline{2-6}
&MSTLF-Frobenius & 8 & 8 & 16/16 & 100$\%$\\
\cline{2-6}
&\textbf{Total} & 64/64 & 64/64\\
\cline{2-4}
&\textbf{\%} & 100$\%$ & 100$\%$\\
\cline{1-4}
\end{tabular}}
\caption{Wilcoxon signed rank test conducted on a per patient basis within \emph{Dataset A} (Mayo simulated data, 8 patients in total) to assess the statistical significance of MSTLF-attention in comparison to other loss function configurations in CycleGAN. Each box shows the number of patients for whom MSTLF-attention is statistically better.}
\label{tab:dataset_A_cycle_un}
\end{table}

\begin{table}[h]
\centering
\resizebox{.7\textwidth}{!}{
\begin{tabular}{|c|c|c|c|c|c|}
\hline
\multicolumn{6}{|c|}{\emph{Dataset A}}\\
\hline
%\multicolumn{2}{c|}{OBJECTIVE FUNCTION}
\multicolumn{2}{|c|}{\textbf{UNIT}} & \textbf{NIQE}& \textbf{PIQUE} & \textbf{Total} & \textbf{\%} \\
\cline{1-6}
\multirow{10}{9em}{MSTLF-attention vs}
&Baseline & 8 & 8 & 16/16 & 100$\%$\\
\cline{2-6}
&VGG-16 & 8 & 8 & 16/16 & 100$\%$\\
\cline{2-6}
&AE-CT  & 7 & 7 & 14/16 & 88$\%$\\
\cline{2-6}
&SSIM-L & 7 & 7 & 14/16 & 88$\%$\\
\cline{2-6}
&EDGE & 7 & 7 & 14/16 & 88$\%$\\
\cline{2-6}
&MSTLF-max & 8 & 8 & 16/16 & 100$\%$\\
\cline{2-6}
&MSTLF-average & 8 & 8 & 16/16 & 100$\%$\\
\cline{2-6}
&MSTLF-Frobenius & 8 & 8 & 16/16 & 100$\%$\\
\cline{2-6}
&\textbf{Total} & 61/64 & 61/64\\
\cline{2-4}
&\textbf{\%} & 95$\%$ & 95$\%$\\
\cline{1-4}
\end{tabular}}
\caption{Wilcoxon signed rank test conducted on a per patient basis within \emph{Dataset A} (Mayo simulated data, 8 patients in total) to assess the statistical significance of MSTLF-attention in comparison to other loss function configurations in UNIT. Each box shows the number of patients for whom MSTLF-attention is statistically better.}
\label{tab:dataset_A_UNIT_un}
\end{table}

\begin{table}[h]
\centering
\resizebox{.7\textwidth}{!}{
\begin{tabular}{|c|c|c|c|c|c|}
\hline
\multicolumn{6}{|c|}{\emph{Dataset B}}\\
\hline
\multicolumn{2}{|c|}{\textbf{Pix2Pix}} & \textbf{NIQE}& \textbf{PIQUE} & \textbf{Total} & \textbf{\%} \\
\cline{1-6}
\multirow{10}{9em}{MSTLF-attention vs}
&Baseline & 8 & 8 & 16/16 & 100$\%$\\
\cline{2-6}
&VGG-16 & 7 & 8 & 15/16 & 94$\%$\\
\cline{2-6}
&AE-CT  & 5 & 5 & 10/16 & 63$\%$\\
\cline{2-6}
&SSIM-L & 5 & 6 & 11/16 & 69$\%$\\
\cline{2-6}
&EDGE & 7 & 8 & 15/16 & 94$\%$\\
\cline{2-6}
&MSTLF-max & 7 & 8 & 15/16 & 94$\%$\\
\cline{2-6}
&MSTLF-average & 6 & 7 & 13/16 & 81$\%$\\
\cline{2-6}
&MSTLF-Frobenius & 6 & 6 & 12/16 & 75$\%$\\
\cline{2-6}
&\textbf{Total} & 51/64 & 56/64\\
\cline{2-4}
&\textbf{\%} & 80$\%$ & 88$\%$\\
\cline{1-4}
\end{tabular}}
\caption{Wilcoxon signed rank test conducted on a per patient basis within \emph{Dataset B} (LIDC/IDRI dataset, 8 patients in total) to assess the statistical significance of MSTLF-attention in comparison to other loss function configurations in Pix2Pix. Each box shows the number of patients for whom MSTLF-attention is statistically better.}
\label{tab:dataset_B_pix2pix}
\end{table}

\begin{table}[h]
\centering
\resizebox{.7\textwidth}{!}{
\begin{tabular}{|c|c|c|c|c|c|}
\hline
\multicolumn{6}{|c|}{\emph{Dataset B}}\\
\hline
\multicolumn{2}{|c|}{\textbf{cycleGAN}} & \textbf{NIQE}& \textbf{PIQUE} & \textbf{Total} & \textbf{\%} \\
\cline{1-6}
\multirow{10}{9em}{MSTLF-attention vs}
&Baseline & 8 & 8 & 16/16 & 100$\%$\\
\cline{2-6}
&VGG-16 & 8 & 8 & 16/16 & 100$\%$\\
\cline{2-6}
&AE-CT  & 8 & 8 & 16/16 & 100$\%$\\
\cline{2-6}
&SSIM-L & 8 & 8 & 16/16 & 100$\%$\\
\cline{2-6}
&EDGE & 8 & 8 & 16/16 & 100$\%$\\
\cline{2-6}
&MSTLF-max & 8 & 8 & 16/16 & 100$\%$\\
\cline{2-6}
&MSTLF-average & 8 & 8 & 16/16 & 100$\%$\\
\cline{2-6}
&MSTLF-Frobenius & 8 & 8 & 16/16 & 100$\%$\\
\cline{2-6}
&\textbf{Total} & 64/64 & 64/64\\
\cline{2-4}
&\textbf{\%} & 100$\%$ & 100$\%$\\
\cline{1-4}
\end{tabular}}
\caption{Wilcoxon signed rank test conducted on a per patient basis within \emph{Dataset B} (LIDC/IDRI dataset, 8 patients in total) to assess the statistical significance of MSTLF-attention in comparison to other loss function configurations in CycleGAN. Each box shows the number of patients for whom MSTLF-attention is statistically better.}
\label{tab:dataset_B_cycle}
\end{table}

\begin{table}[h]
\centering
\resizebox{.7\textwidth}{!}{
\begin{tabular}{|c|c|c|c|c|c|}
\hline
\multicolumn{6}{|c|}{\emph{Dataset B}}\\
\hline
\multicolumn{2}{|c|}{\textbf{UNIT}} & \textbf{NIQE}& \textbf{PIQUE} & \textbf{Total} & \textbf{\%} \\
\cline{1-6}
\multirow{10}{9em}{MSTLF-attention vs}
&Baseline & 8 & 7 & 16/16 & 94$\%$\\
\cline{2-6}
&VGG-16 & 8 & 8 & 16/16 & 100$\%$\\
\cline{2-6}
&AE-CT  & 8 & 6 & 16/16 & 88$\%$\\
\cline{2-6}
&SSIM-L & 8 & 5 & 16/16 & 81$\%$\\
\cline{2-6}
&EDGE & 8 & 0 & 16/16 & 50$\%$\\
\cline{2-6}
&MSTLF-max & 8 & 2 & 16/16 & 63$\%$\\
\cline{2-6}
&MSTLF-average & 8 & 7 & 16/16 & 94$\%$\\
\cline{2-6}
&MSTLF-Frobenius & 8 & 7 & 16/16 & 94$\%$\\
\cline{2-6}
&\textbf{Total} & 64/64 & 42/64\\
\cline{2-4}
&\textbf{\%} & 100$\%$ & 67$\%$\\
\cline{1-4}
\end{tabular}}
\caption{Wilcoxon signed rank test conducted on a per patient basis within \emph{Dataset B} (LIDC/IDRI dataset, 8 patients in total) to assess the statistical significance of MSTLF-attention in comparison to other loss function configurations in UNIT. Each box shows the number of patients for whom MSTLF-attention is statistically better.}
\label{tab:dataset_B_UNIT}
\end{table}

\begin{table}[h]
\centering
\resizebox{.7\textwidth}{!}{
\begin{tabular}{|c|c|c|c|c|c|}
\hline
\multicolumn{6}{|c|}{\emph{Dataset C}}\\
\hline
\multicolumn{2}{|c|}{\textbf{Pix2Pix}} & \textbf{NIQE}& \textbf{PIQUE} & \textbf{Total} & \textbf{\%} \\
\cline{1-6}
\multirow{10}{9em}{MSTLF-attention vs}
&Baseline & 44 & 31 & 75/100 & 75$\%$\\
\cline{2-6}
&VGG-16 & 47 & 24 & 71/100 & 71$\%$\\
\cline{2-6}
&AE-CT  & 21 & 12 & 33/100 & 33$\%$\\
\cline{2-6}
&SSIM-L & 26 & 7 & 33/100 & 33$\%$\\
\cline{2-6}
&EDGE & 50 & 38 & 88/100 & 88$\%$\\
\cline{2-6}
&MSTLF-max & 43 & 27 & 70/100 & 70$\%$\\
\cline{2-6}
&MSTLF-average & 36 & 16 & 52/100 & 52$\%$\\
\cline{2-6}
&MSTLF-Frobenius & 26 & 18 & 44/100 & 44$\%$\\
\cline{2-6}
&\textbf{Total} & 293/400 & 173/400\\
\cline{2-4}
&\textbf{\%} & 73$\%$ & 43$\%$\\
\cline{1-4}
\end{tabular}}
\caption{Wilcoxon signed rank test conducted on a per patient basis within \emph{Dataset C} (ELCAP dataset, 50 patients in total) to assess the statistical significance of MSTLF-attention in comparison to other loss function configurations in Pix2Pix. Each box shows the number of patients for whom MSTLF-attention is statistically better.}
\label{tab:dataset_C_pix2pix}
\end{table}

\begin{table}[h]
\centering
\resizebox{.7\textwidth}{!}{
\begin{tabular}{|c|c|c|c|c|c|}
\hline
\multicolumn{6}{|c|}{\emph{Dataset C}}\\
\hline
\multicolumn{2}{|c|}{\textbf{cycleGAN}} & \textbf{NIQE}& \textbf{PIQUE} & \textbf{Total} & \textbf{\%} \\
\cline{1-6}
\multirow{10}{9em}{MSTLF-attention vs}
& Baseline & 50 & 50 & 100/100 & 100$\%$\\
\cline{2-6}
& VGG-16 & 50 & 49 & 99/100 & 99$\%$\\
\cline{2-6}
& AE-CT  & 50 & 50 & 100/100 & 100$\%$\\
\cline{2-6}
& SSIM-L & 50 & 50 & 100/100 & 100$\%$\\
\cline{2-6}
& EDGE & 50 & 50 & 100/100 & 100$\%$\\
\cline{2-6}
& MSTLF-max & 50 & 50 & 100/100 & 100$\%$\\
\cline{2-6}
& MSTLF-average & 50 & 50 & 100/100 & 100$\%$\\
\cline{2-6}
& MSTLF-Frobenius & 50 & 49 & 99/100 & 99$\%$\\
\cline{2-6}
& \textbf{Total} & 400/400 & 398/400\\
\cline{2-4}
& \textbf{\%} & 100$\%$ & 99$\%$\\
\cline{1-4}
\end{tabular}}
\caption{Wilcoxon signed rank test conducted on a per patient basis within \emph{Dataset C} (ELCAP dataset, 50 patients in total) to assess the statistical significance of MSTLF-attention in comparison to other loss function configurations in CycleGAN. Each box shows the number of patients for whom MSTLF-attention is statistically better.}
\label{tab:dataset_C_cycle}
\end{table}

\begin{table}[t]
\centering
\resizebox{.7\textwidth}{!}{
\begin{tabular}{|c|c|c|c|c|c|}
\hline
\multicolumn{6}{|c|}{\emph{Dataset C}}\\
\hline
\multicolumn{2}{|c|}{\textbf{UNIT}} & \textbf{NIQE}& \textbf{PIQUE} & \textbf{Total} & \textbf{\%} \\
\cline{1-6}
\multirow{10}{9em}{MSTLF-attention vs}
& Baseline & 50 & 38 & 88/100 & 88$\%$\\
\cline{2-6}
& VGG-16 & 50 & 47 & 97/100 & 97$\%$\\
\cline{2-6}
& AE-CT  & 50 & 22 & 72/100 & 72$\%$\\
\cline{2-6}
& SSIM-L & 50 & 12 & 62/100 & 62$\%$\\
\cline{2-6}
& EDGE & 50 & 0 & 50/100 & 50$\%$\\
\cline{2-6}
& MSTLF-max & 50 & 7 & 57/100 & 57$\%$\\
\cline{2-6}
& MSTLF-average & 50 & 43 & 93/100 & 93$\%$\\
\cline{2-6}
& MSTLF-Frobenius & 50 & 34 & 84/100 & 84$\%$\\
\cline{2-6}
& \textbf{Total} & 400/400 & 203/400\\
\cline{2-4}
& \textbf{\%} & 100$\%$ & 51$\%$\\
\cline{1-4}
\end{tabular}}
\caption{Wilcoxon signed rank test conducted on a per patient basis within \emph{Dataset C} (ELCAP dataset, 50 patients in total) to assess the statistical significance of MSTLF-attention in comparison to other loss function configurations in UNIT. Each box shows the number of patients for whom MSTLF-attention is statistically better.}
\label{tab:dataset_C_UNIT}
\end{table}

%%%%%%%%%%%%%%%%%%%%%%%%%%%%%%%%%%%%%%%%%%%%%%%%%%%%%%%%%%%%%%%%%%%%%%%%%%%%%
%%%%%%%%%%%%%%%%%%%%%%%%%%%%%%%%%%%%%%%%%%%%%%%%%%%%%%%%%%%%%%%%%%%%%%%%%%%%%
\FloatBarrier
\section{Visual results} \label{suppl:visuals}
\begin{figure}[b!]
\centering
\includegraphics[width=9cm]{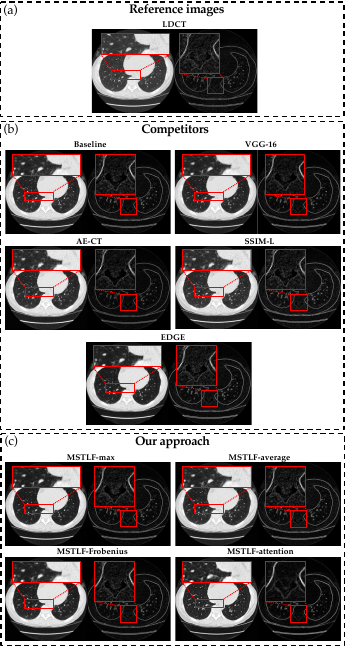}
\caption{
Visual comparison of denoised CT slices from the LIDC/IDRI dataset ({\em Dataset B}). 
We show all loss function configurations for CycleGAN.
The images are organized into ``Reference Images'' (panel(a)), ``Competitors'' (panel (b)) and ``Our Approach'' (panel (c)). For each configuration, results are displayed in the image domain (left image, with a display window of [-1200, 200] HU) and the gradient domain (right image).
Zoomed ROIs highlight key regions demonstrating noise reduction effectiveness.
}
\label{fig:cycleGAN_test_3}
\end{figure}

\begin{figure}[h!]
\centering
\includegraphics[width=9cm]{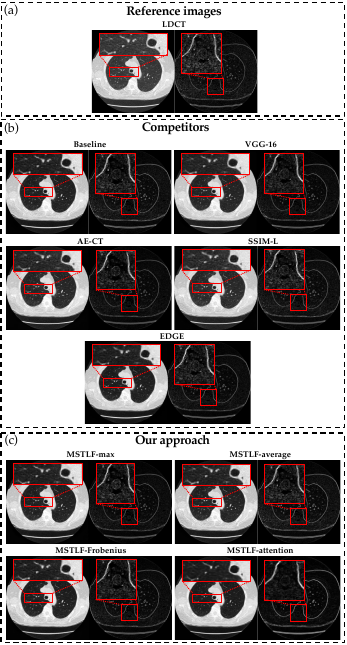}
\caption{
Visual comparison of denoised CT slices from the ELCAP dataset ({\em Dataset C}). 
We show all loss function configurations for CycleGAN.
The images are organized into ``Reference Images'' (panel(a)), ``Competitors'' (panel (b)) and ``Our Approach'' (panel (c)). For each configuration, results are displayed in the image domain (left image, with a display window of [-1200, 200] HU) and the gradient domain (right image).
Zoomed ROIs highlight key regions demonstrating noise reduction effectiveness.
}
\label{fig:cycleGAN_test_4}
\end{figure}

\begin{figure}[h!]
\centering
\includegraphics[width=9cm]{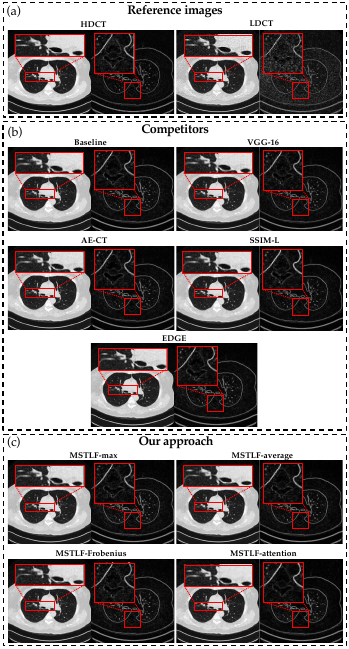}
\caption{
Visual comparison of denoised CT slices from the Mayo Clinic simulated data ({\em Dataset A}). 
We show all loss function configurations for Pix2Pix.
The images are organized into ``Reference Images'' (panel(a)), ``Competitors'' (panel (b)) and ``Our Approach'' (panel (c)). For each configuration, results are displayed in the image domain (left image, with a display window of [-1200, 200] HU) and the gradient domain (right image).
Zoomed ROIs highlight key regions demonstrating noise reduction effectiveness.
}
\label{fig:pix2pix_test_2}
\end{figure}

\begin{figure}[h!]
\centering
\includegraphics[width=9cm]{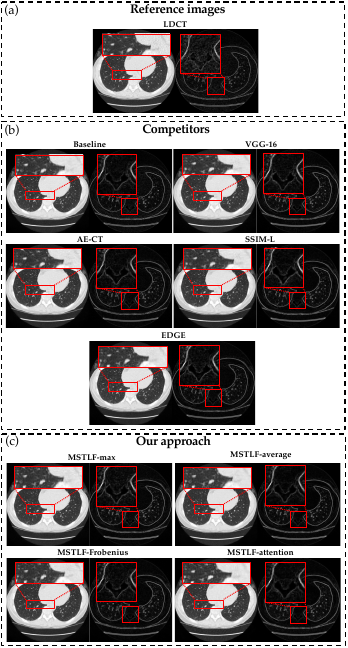}
\caption{
Visual comparison of denoised CT slices from the LIDC/IDRI dataset ({\em Dataset B}). 
We show all loss function configurations for Pix2Pix. 
The images are organized into ``Reference Images'' (panel(a)), ``Competitors'' (panel (b)) and ``Our Approach'' (panel (c)). For each configuration, results are displayed in the image domain (left image, with a display window of [-1200, 200] HU) and the gradient domain (right image).
Zoomed ROIs highlight key regions demonstrating noise reduction effectiveness.
}
\label{fig:pix2pix_test_3}
\end{figure}

\begin{figure}[b!]
\centering
\includegraphics[width=9cm]{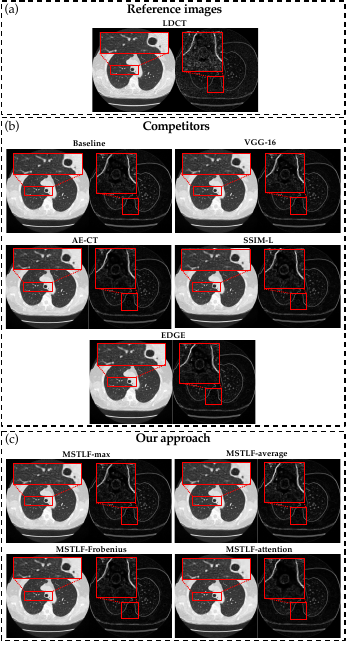}
\caption{
Visual comparison of denoised CT slices from the ELCAP dataset ({\em Dataset C}). 
We show all loss function configurations for Pix2Pix. 
The images are organized into ``Reference Images'' (panel(a)), ``Competitors'' (panel (b)) and ``Our Approach'' (panel (c)). For each configuration, results are displayed in the image domain (left image, with a display window of [-1200, 200] HU) and the gradient domain (right image).
Zoomed ROIs highlight key regions demonstrating noise reduction effectiveness.
}
\label{fig:pix2pix_test_4}
\end{figure}

\begin{figure}[b!]
\centering
\includegraphics[width=9cm]{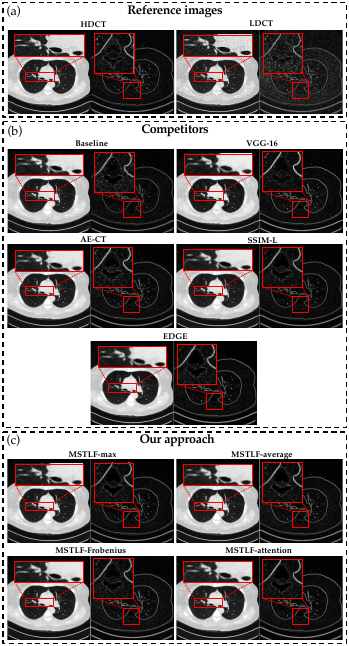}
\caption{
Visual comparison of denoised CT slices from the Mayo Clinic simulated data ({\em Dataset A}). 
We show all loss function configurations for UNIT. 
The images are organized into ``Reference Images'' (panel(a)), ``Competitors'' (panel (b)) and ``Our Approach'' (panel (c)). For each configuration, results are displayed in the image domain (left image, with a display window of [-1200, 200] HU) and the gradient domain (right image).
Zoomed ROIs highlight key regions demonstrating noise reduction effectiveness.
}
\label{fig:unit_test_2}
\end{figure}

\begin{figure}[b!]
\centering
\includegraphics[width=9cm]{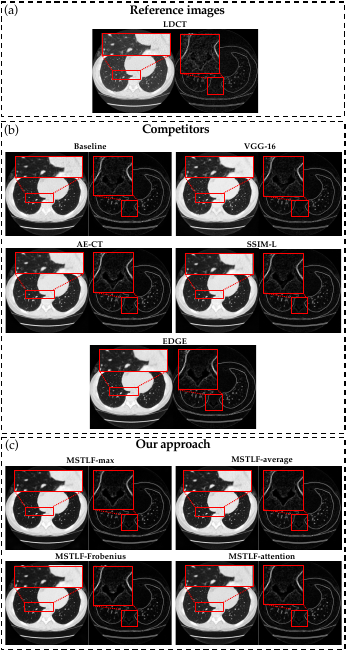}
\caption{
Visual comparison of denoised CT slices from the LIDC/IDRI dataset ({\em Dataset B}). 
We show all loss function configurations for UNIT. 
The images are organized into ``Reference Images'' (panel(a)), ``Competitors'' (panel (b)) and ``Our Approach'' (panel (c)). For each configuration, results are displayed in the image domain (left image, with a display window of [-1200, 200] HU) and the gradient domain (right image).
Zoomed ROIs highlight key regions demonstrating noise reduction effectiveness.
}
\label{fig:unit_test_3}
\end{figure}

\begin{figure}[b!]
\centering
\includegraphics[width=9cm]{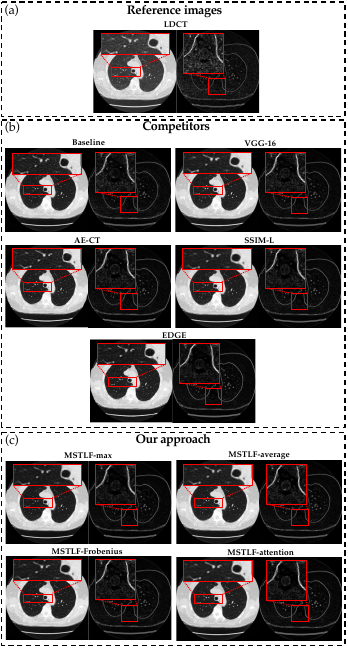}
\caption{
Visual comparison of denoised CT slices from the ELCAP dataset ({\em Dataset C}). 
We show all loss function configurations for UNIT. 
The images are organized into ``Reference Images'' (panel(a)), ``Competitors'' (panel (b)) and ``Our Approach'' (panel (c)). For each configuration, results are displayed in the image domain (left image, with a display window of [-1200, 200] HU) and the gradient domain (right image).
Zoomed ROIs highlight key regions demonstrating noise reduction effectiveness.
}
\label{fig:unit_test_4}
\end{figure}

This section presents visual results attained using CycleGAN, Pix2Pix, and UNIT across different configurations of the loss function and different datasets. 
 The images are organized into three panels: panel (a) shows ``Reference Images'', panel (b) shows ``Competitors'', and pabel (c) shows ``Our Approach''. For each configuration, results are displayed in the image domain (left image, with a display window of [-1200, 200] HU) and the gradient domain (right image).
Zoomed ROIs highlight key regions demonstrating noise reduction effectiveness.
\figurename~\ref{fig:cycleGAN_test_3} and \ref{fig:cycleGAN_test_4} depict examples of denoised images obtained using CycleGAN applied to \emph{Dataset B} (LIDC/IDRI dataset) and \emph{Dataset C} (ELCAP dataset), respectively.  \figurename~\ref{fig:pix2pix_test_2}, \ref{fig:pix2pix_test_3}, and \ref{fig:pix2pix_test_4} show examples of denoised images obtained through Pix2Pix on \emph{Dataset A} (Mayo simulated data), \emph{Dataset B} (LIDC/IDRI dataset), and \emph{Dataset C} (ELCAP dataset), respectively. Furthermore, \figurename~\ref{fig:unit_test_2}, \ref{fig:unit_test_3}, and \ref{fig:unit_test_4} show examples of denoised images produced by UNIT on \emph{Dataset A} (Mayo simulated data), \emph{Dataset B} (LIDC/IDRI dataset), and \emph{Dataset C} (ELCAP dataset), respectively.
As mentioned in the main manuscript, MSTLF-attention produced the image with the lowest gradient that, in turn, corresponds to the lowest level of noise while preserving the lung boundaries, as shown in  \figurename~\ref{fig:cycleGAN_test_3},~\ref{fig:cycleGAN_test_4},~\ref{fig:pix2pix_test_2},~\ref{fig:pix2pix_test_3},~\ref{fig:pix2pix_test_4},~\ref{fig:unit_test_2},~\ref{fig:unit_test_3},~\ref{fig:unit_test_4} (c).

\FloatBarrier
\bibliographystyle{model2-names.bst}\biboptions{authoryear}
\bibliography{refs}